\documentclass[%
reprint,
superscriptaddress,
amsmath,amssymb,
aps,
]{revtex4-2}

\usepackage{graphicx}
\usepackage{dcolumn}
\usepackage{bm}
\usepackage{hyperref}
\usepackage{siunitx}
\usepackage{color}
\usepackage{placeins}
\usepackage{overpic}
\definecolor{IB}{rgb}{0.17, 0.50, 0.72}
\definecolor{DL}{rgb}{0.25, 0.5, 1}
\definecolor{SJ}{rgb}{0.25, 0.5, 0.5}
\definecolor{JG}{rgb}{0.85, 0.25, 0.85}
\DeclareSIUnit\px{px}

\begin{document}
	
	\preprint{APS/123-QED}
	
	\title{Probing the fluctuating magnetic field of Fe-triazole spin-crossover thin-layers with nitrogen-vacancy centers in diamond}
	
	\author{Isabel Cardoso Barbosa}
     \affiliation{
		Department of Physics and State Research Center OPTIMAS, University of Kaiserslautern-Landau, Erwin-Schroedinger-Str. 46,
        67663 Kaiserslautern, Germany\\
	}
	\author{Tim Hochd{\"o}rffer}%
    \affiliation{
		Department of Biophysics and Medical Physics, University of Kaiserslautern-Landau, Erwin-Schroedinger-Str. 52,
        67663 Kaiserslautern, Germany\\
	}
	\author{Juliusz A. Wolny}%
    \affiliation{
		Department of Biophysics and Medical Physics, University of Kaiserslautern-Landau, Erwin-Schroedinger-Str. 52,
        67663 Kaiserslautern, Germany\\
	}
	\author{Dennis L{\"o}nard}%
     \affiliation{
		Department of Physics and State Research Center OPTIMAS, University of Kaiserslautern-Landau, Erwin-Schroedinger-Str. 46,
        67663 Kaiserslautern, Germany\\
	}
	\author{Stefan Johansson}%
     \affiliation{
		Department of Physics and State Research Center OPTIMAS, University of Kaiserslautern-Landau, Erwin-Schroedinger-Str. 46,
        67663 Kaiserslautern, Germany\\
	}
	\author{Jonas Gutsche}%
     \affiliation{
		Department of Physics and State Research Center OPTIMAS, University of Kaiserslautern-Landau, Erwin-Schroedinger-Str. 46,
        67663 Kaiserslautern, Germany\\
	}
	\author{Volker Sch{\"u}nemann}%
    \affiliation{
		Department of Biophysics and Medical Physics, University of Kaiserslautern-Landau, Erwin-Schroedinger-Str. 52,
        67663 Kaiserslautern, Germany\\
	}
	\author{Artur Widera}%
     \affiliation{
		Department of Physics and State Research Center OPTIMAS, University of Kaiserslautern-Landau, Erwin-Schroedinger-Str. 46,
        67663 Kaiserslautern, Germany\\
	}
    \email{Author to whom correspondence should be addressed: widera@physik.uni-kl.de}

	\date{\today}

	\begin{abstract}
        Fe$^{\mathrm{II}}$ spin-crossover (SCO) complexes are materials that change their magnetic properties upon temperature variation, exhibiting a thermal hysteresis.
        Particularly interesting for magnetic-memory applications are thin layers of SCO complexes, where practical magnetic probing techniques are required.
        While conventional magnetometry on SCO complexes employs cryogenic temperatures, nitrogen-vacancy (NV) centers are quantum magnetometers that can operate at room temperature with high spatial resolution and magnetic-field sensitivity.
        In this work, we apply thin layers of Fe-triazole SCO complexes directly onto a single-crystal diamond with shallow NV centers working as magnetic sensors and probe the fluctuating magnetic field. 
        Using temperature-dependent NV-center $T_1$ measurements and a widefield technique, we find that the complexes are paramagnetic in the investigated temperature range from \SI{20}{\degreeCelsius} to \SI{80}{\degreeCelsius}.
        We quantitatively describe the $T_1$ time by a model considering the fluctuating magnetic field of the Fe$^{\mathrm{II}}$ ions.
        We see signatures of a local change of spin state in the $T_1$ relaxometry data, but structural changes in the SCO material dominate the local magnetic environment of the NV centers.
        Moreover, we conduct a Hahn echo to measure the $T_2$ time, which contrasts the findings of the $T_1$ times for the SCO complexes. 
        We attribute this to different NV detection sensitivities towards Fe$^{\mathrm{II}}$ and Fe$^{\mathrm{III}}$ of the protocols.
        Our results on the magnetic properties of SCO materials highlight the capabilities of the NV center as a susceptible sensor for fluctuating magnetic fields. At the same time, a spin switching of the complexes cannot be observed due to the systematic challenges when working on nanometer distances to the SCO thin layers.
	
	\end{abstract}
	
	\maketitle
		\section{\label{sec:level1}INTRODUCTION}
        NV centers in diamonds provide a spin-1 system susceptible to noise originating from fluctuating magnetic fields, influencing the NV centers' longitudinal relaxation time $T_1$ and the dephasing time $T_2$ \cite{Levine.2019, Steinert.2013, SchaferNolte.2014}.
        Relaxometry is an established method of measuring the NV centers' $T_1$ time to detect high-frequency fluctuating magnetic fields from paramagnetic ions or molecules \cite{Levine.2019, Mzyk.2022}.
        The effect of Gd$^{\mathrm{III}}$ ions in aqueous solutions has been studied \cite{Iyer.20240130}, and from relaxation measurements, the concentration of Gd$^{\mathrm{III}}$ ions has been determined \cite{Steinert.2013}.
        In addition, this method has been shown to detect changes in the pH in aqueous solutions \cite{Fujisaku.2019, Cheng.2024} or to follow the process of chemical reactions \cite{PeronaMartinez.2020, Li.2022}.
        In the context of biological applications, relaxometry has been used to examine biomolecules of Fe$^{\mathrm{III}}$, such as ferritin \cite{SchaferNolte.2014, Grant.2023}, methemoglobin \cite{Lamichhane.2024}, and cytochrome C \cite{Lamichhane.2024_2}.
        Also, the paramagnetic Fe$^{\mathrm{II}}$ ion in hemoglobin has been shown to reduce the NV centers' $T_1$ time \cite{Gorrini.2019}.
        Complementary to the $T_1$ time, the $T_2$ time of NV centers is influenced by low-frequency magnetic noise in superparamagnets \cite{SchaferNolte.2014, SchmidLorch.2015} and paramagnets \cite{Steinert.2013}.
        Another system exhibiting interesting magnetic properties are iron-based SCO complexes that can change their spin state upon changing the temperature \cite{Gutlich.2000}.
        In detail, Fe$^{\mathrm{II}}$ SCO complexes switch from a diamagnetic low-spin (LS) state of spin quantum number $S=0$ to a paramagnetic high-spin (HS) state of $S=2$ with a thermal hysteresis \cite{Roubeau.2012, Brooker.2015}.
        For the application of memory devices, thin-layer samples of SCO complexes are especially of interest \cite{Hochdorffer.2019}.
        The magnetic properties of these complexes are usually examined using a Superconducting Quantum Interference Device (SQUID) \cite{Hogue.2018}, which operates at cryogenic temperatures \cite{Bao.2023, BishopVanHorn.2019}.
        Recent emerging magnetometers are based on diamond NV centers, which allow the detection of static, alternating, and fluctuating magnetic fields with high sensitivity and spatial resolution at room temperature \cite{Barry.2020, Hall.2009}.
        The capabilities of NV centers in combination with complexes that switch their magnetic properties have recently been demonstrated for SCO nanorods \cite{Lamichhane.2023} and SCO metal-organic frameworks \cite{Flinn.2024} using optically detected magnetic resonance (ODMR).
        
\begin{figure}[]
         \begin{overpic}[width=86mm]{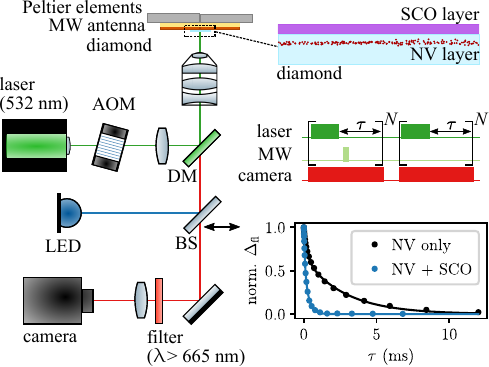}
             \put(-1, 72.5){(a)}
             \put(49,52){(b)}
             \put(49,30){(c)}
            \end{overpic}
                \caption{\label{fig:setup}
                Schematics of $T_1$ measurements in widefield configuration. (a) Microscope setup. 
                A laser beam of wavelength \SI{532}{\nano\meter} is pulsed by an acousto-optic modulator (AOM) and focused on the objective's back focal plane to homogeneously illuminate a large section of the NV diamond sample. 
                NV fluorescence is filtered by a 550-nm dichroic mirror (DM) and a 665-nm longpass filter and focused onto a camera.
                For orientation on the sample, a beam splitter (BS) can be moved into the beam path to illuminate the sample with light from a light-emitting diode (LED).
                The diamond sample has shallow NV centers implanted a few nm beneath its surface. SCO samples are directly drop-cast on the diamond. 
                A microwave (MW) antenna guides MW pulses to the NV centers. 
                The sample's temperature is regulated with Peltier elements.
                (b) Pulsed sequence used for relaxometry measurements. 
                (c) $T_1$ decays for the NV centers in the clean diamond ($T_1^{\prime} = \SI{2.80 \pm 0.12}{\milli\second}$) and an example with SCO complexes applied (\textbf{SCO~I}, $T_1 = \SI{0.27 \pm 0.01}{\milli\second}$). 
                Solid lines are biexponential fit curves to the data points described in the main text.
                } 
        \end{figure}

        This work employs a shallow NV layer in diamond to probe two Fe-triazole SCO complexes in thin-layer form, see Fig.~\ref{fig:setup}(a).
        The complexes we investigate here are $[\mathrm{Fe}(\mathrm{Htrz})_2(\mathrm{trz})](\mathrm{BF}_{4})$, termed \textbf{SCO~I}, and $[\mathrm{Fe}(\mathrm{atrz})_3](\mathrm{CH}_3\mathrm{SO}_3)_{4/3}(\mathrm{SO}_4)_{1/3}$, termed \textbf{SCO~II} (Htrz = 1,2,4-4H-triazole, trz = 1,2,4-triazolate, atrz = 4-amino-1,2,4-triazole).
        These 1D-polymeric complexes exhibit spin-switching temperatures above room temperature \cite{Kroeber.1994, Roubeau.2012, Hochdorffer.2019, Hochdorffer.2023}.
        We use temperature-dependent NV-center relaxometry of the sequence depicted in Fig.~\ref{fig:setup}(b) to investigate the fluctuating magnetic field originating from complexes \textbf{SCO~I} and \textbf{SCO~II} with spatial resolution.
        In comparison to the clean diamond, we observe a decrease of the NV-center $T_1$ time with SCO complexes applied on the diamond, see Fig.~\ref{fig:setup}(c).
        Modeling the HS SCO layer on the diamond, we derive an estimated $T_1$ time for an NV center and compare the calculation to our experimental results. 
        We compare the results for the $T_1$ times at different temperatures to the spin state of the SCO complexes we obtain with conventional Raman spectroscopy.
        In addition to that, we probe \textbf{SCO~II} with a Hahn-echo measurement to obtain $T_2$ times of the NV centers close to the SCO layer. 
        Since the $T_1$ time of the NV centers itself exhibits a temperature dependence \cite{Jarmola.2012, Norambuena.2018}, we analyze the $T_1$ times for the NV-SCO samples at the same temperature in the heating and cooling branches of the hystereses. 
        
	\section{\label{sec:methods}SAMPLE PREPARATION}
    
    \subsection{Diamond sample preparation}
	    An ELSC (Single Crystal Electronic)-grade diamond chip of size \qtyproduct[product-units=single]{2 x 2 x 0.5}{\milli\meter\cubed} from Element Six Technologies was irradiated with $^{14}\mathrm{N}^{+}$ at an energy of \SI{6}{\kilo \electronvolt} and a dose of \SI{4e13}{\per \centi \meter \squared} at an angle of \SI{7}{\degree}. 
        Using a simulation \cite{SRIM}, we estimate the depth of the NV layer to be $\bar{d} = \SI{9.3}{\nano \meter}$ of width $\sigma = \SI{3.6}{\nano \meter}$ in a Gaussian distribution.
	    The diamond was then annealed in vacuum ($\SI{<7.8e-7}{\milli \bar}$), and the temperature was successively held at \SI{400}{\degreeCelsius} (\SI{4}{\hour}) and \SI{800}{\degreeCelsius} (\SI{2}{\hour}) \cite{Chu.2014, Appel.2016} to enhance the yield of NV centers. 
	    Afterward, the diamond chip was treated in a tri-acid solution (equal ratios of sulfuric acid (\SI{96}{\percent}), perchloric acid (\SI{70}{\percent}), and nitric acid (\SI{65}{\percent})) and heated to \SI{500}{\degreeCelsius} (\SI{1}{\hour}) \cite{Chu.2014, Appel.2016} for diamond surface oxidation.
	    Finally, the diamond was cleaned in acetone and isopropyl alcohol. 
     
        We perform $T_1$ measurements with the clean diamond chip or with SCO complexes directly cast on the diamond chip. 
        To change the SCO sample on the diamond, SCO samples are removed from the diamond chip with alcohol/water solutions in an ultrasonic bath. 
	    We observe a recovery of the $T_1$ time from short $T_1$ times with the SCO complexes applied to long $T_1$ times after the cleaning process.
        Without any SCO complexes applied, we determine $T_1^{\prime} = \SI{2.80 \pm 0.12}{\milli \second}$. 
        For a study on the temperature dependence of the $T_1$ time in our sample, see Supporting Information.
	    
	\subsection{SCO synthesis}
    \textbf{SCO I} was synthesized according to the method described in Ref.~\cite{Kroeber.1994}.
    \textbf{SCO II} was synthesized in a similar way, using a 2:1 mixture of $\mathrm{Fe}(\mathrm{CH}_3\mathrm{SO}_3)_2 \cdot 7 \mathrm{H}_2\mathrm{O}$ and $\mathrm{Fe}\mathrm{SO}_4 \cdot 7 \mathrm{H}_2\mathrm{O}$.
    The thin films were prepared as described below, similar to the method described in Ref.~\cite{Hochdorffer.2019}.
	
	\subsection{SCO application on diamond}
	We dissolved the SCO complexes \textbf{SCO~I} and \textbf{SCO~II} in water/alcohol solutions and added ascorbic acid to all solutions to prevent the SCO compounds from oxidizing. 
	This resulted in colorless and transparent solutions.
 
	    \subsubsection*{\textbf{SCO~I}}
	    We dissolved \SI{45}{\milli \gram} of compound \textbf{SCO~I} in \SI{4}{\milli \liter} water and \SI{1}{\milli \liter} ethanol. 
	    We drop-cast $\SI{\approx 1}{\micro \liter}$ on the diamond chip and removed the solvent on a hot plate at $\SI{\approx 80}{\degreeCelsius}$.
        This resulted in an amount of $n \approx \SI{0.026}{\micro \mol}$ of SCO material on the diamond chip.
	    The LS state of the SCO complexes was prepared by letting the sample cool down to room temperature. 
        
	    Since the temperature for reaching the complexes' HS state (\SI{110}{\degreeCelsius}) lay beyond the capabilities of our sample holder (max. \SI{80}{\degreeCelsius}) for NV measurements, we applied external heating to the SCO-diamond sample. 
	    We first heated the sample holder to its maximum temperature and then put the sample on a hot plate to reach \SI{115}{\degreeCelsius}. 
	    Next, we removed the external hot plate and held the temperature at \SI{80}{\degreeCelsius} to stay on the cooling branch of the hysteresis.
     
        \subsubsection*{\textbf{SCO~II}}
	    Two similar solutions of complex \textbf{SCO~II} were used for the measurements in this paper.
        We prepared solution 1 by adding \SI{64}{\milli \gram} of \textbf{SCO~II} to \SI{5}{\milli \liter} ethanol and \SI{2}{\milli \liter} water.
	    For sample 1, we drop-cast $\SI{\approx 1}{\micro \liter}$ of solution 1 on the diamond chip. 
        We removed the solvent on a hot plate at $\SI{\approx 60}{\degreeCelsius}$, which resulted in an amount of SCO material applied of $n \approx \SI{0.020}{\micro \mol}$.
	    Solution 2 was prepared by dissolving \SI{40}{\milli \gram} of \textbf{SCO~II} in \SI{5}{\milli \liter} water and \SI{2}{\milli \liter} methanol. 
	    For sample 2, we drop-cast $\SI{\approx 3}{\micro \liter}$ of solution 2 on the diamond chip.
	    This was done by application of drops of volume $\SI{\approx 1}{\micro \liter}$, removing the solvent, and repeating the process.
	    The solvent was removed on a hot plate at $\SI{\approx 60}{\degreeCelsius}$.
        This process resulted in an amount of SCO material applied of $n \approx \SI{0.037}{\micro \mol}$.
        
	    To transfer samples of \textbf{SCO~II} to the LS state, the colorless samples were cooled to \SI{-20}{\degreeCelsius} for \SI{15}{\minute} to \SI{20}{\minute}. 
	    This resulted in a color change of the thin-layer samples from colorless to purple, indicating that the sample switched to the LS state \cite{Brooker.2015}.
     
     \subsection{Raman spectroscopy}
     We recorded Raman spectra of the SCO thin-layer samples on the diamond substrate.
     Raman experiments were performed with a Senterra© Raman Microscope (Bruker) using an excitation wavelength of \SI{532}{\nano \meter}, at a laser power of \SI{2}{\milli \watt}. 
     The NV diamonds which were coated with the SCO materials were positioned inside a minicryostat (Linkam© stage FTIR600) and measurements were conducted in the temperature range of \SI{290}{\kelvin} to \SI{400}{\kelvin}. 
     Local heating due to laser illumination is estimated to be \SI{\approx 10}{\kelvin} at the irradiated sample spot (diameter \SI{<2}{\micro\meter} to \SI{3}{\micro\meter}). 
     Successive cycles of the SCO transition were investigated by cooling the NV diamonds to low temperatures within a liquid N$_2$ atmosphere or cooling to \SI{-20}{\degreeCelsius} in the cooler between cycles. 

    \section{NV-SENSING METHODS}
    
	\subsection{$T_1$ measurements}
    	We measure the NV centers' $T_1$ time in a microscope setup, as shown in Fig.~\ref{fig:setup}(a). 
    	The NV centers are excited with a 532-nm laser of power of $\SI{\approx 80}{\milli \watt}$. 
    	Laser pulses are created with an acousto-optic modulator (AOM).
    	The laser light is focused onto the objective's (Nikon MRH08430, \SI{40}{\times}/\SI{0.60}{}) back focal plane to illuminate a large area of the NV layer in the diamond chip of $\SI{\approx 140}{\micro \meter}$ diameter. 
    	Using a 550-nm dichroic longpass, the NV fluorescence is filtered. 
    	Additionally, we specifically filter the NV$^{-}$ fluorescence using a 665-nm longpass filter. 
    	The NV fluorescence is detected with an electron-multiplying charge coupled device (EMCCD) camera. 
    	For orientation on the sample before measurements, light from a white-light LED (light-emitting diode) can be guided through the objective with a removable 50:50 beam splitter and illuminate the sample homogeneously. 
    	The diamond chip is mounted in a temperature-controlled sample holder, which allows us to perform measurements between $\SI{20}{\degreeCelsius}$ and $\SI{\approx 80}{\degreeCelsius}$.
    	To drive transitions between the NV centers' spin states $m_s$, a microwave (MW) antenna is mounted directly above the NV diamond chip. 
        We fabricated the antenna according to Ref.~\cite{Sasaki.2016}.
    	The diamond chip is oriented with its NV layer pointing towards the MW antenna to ensure high MW powers at the position of the NV-center layer.
    	We employ Peltier elements to regulate the sample's temperature and a thermistor to read the temperature.
    	
    	We perform relaxometry experiments without an external magnetic field to probe NV centers of all orientations at once, as previously reported in Refs.~\cite{Steinert.2013, Ziem.2013}, since we expect an equal influence of the paramagnetic Fe$^{\mathrm{II}}$ ions independent of the NV-center orientation.
    	The pulsed sequence consists of laser pulses of \SI{50}{\micro \second} duration for NV-center spin polarization and readout, separated by a variable relaxation time $\tau$, see Fig.~\ref{fig:setup}(b).
    	The pulses are repeated $N$ times ($N \approx \SI{2000}{}$), during which the camera is exposed. 
    	In the first part of the measurement, an MW pulse transfers the NV centers from their spin ground state to an excited spin state.
        It follows $\SI{\approx 1}{\micro \second}$ after the laser pulse and lies within $\tau$.
    	The sequence is repeated without the MW pulse being applied, and the camera is exposed again. 
    	For data acquisition, 40 exposures are accumulated for the sequence to one image with and without the MW pulse applied alternatingly for the same value of $\tau$. 
    	The entire sequence is repeated 10 times.
        The $T_1$ signal is obtained by forming the difference between the two images for every value of $\tau$, yielding the fluorescence difference \cite{Jarmola.2012, Steinert.2013}.
    	During the data evaluation, \qtyproduct[product-units=single]{10 x 10}{\px \squared}, corresponding to \qtyproduct[product-units=single]{2 x 2}{\micro \meter \squared} on the diamond surface, are binned together.
        This resolution of $\SI{2}{\micro\meter}$ is in the order of the expected resolution for our setup \cite{Nishimura.20240222} and is sufficient for the investigations on a thin-layer sample.
        A mean fluorescence-difference decay $\Delta_{\mathrm{fl}} (\tau)$ is calculated for every binned pixel. 
    	We normalize each $\Delta_{\mathrm{fl}} (\tau)$ to its maximum and fit a function of type 
    	\begin{equation}
    	    \Delta_{\mathrm{fl}} (\tau) = A \exp(-\tau/T_{1,1}) + B \exp(-\tau/T_{1,2})
    	\end{equation}
    	for every binned pixel.
    	As pointed out in previous works, this biexponential $T_1$ decay is commonly observed in NV centers close to the diamond surface \cite{FreireMoschovitis.2023, Ziem.2013}. 
    	Consistent with previous studies \cite{Steinert.2013, Ziem.2013, PeronaMartinez.2020, FreireMoschovitis.2023}, all $T_1$ times mentioned in this paper refer to the longer $T_1$ component found in the biexponential fit.
        The shorter $T_1$ component is attributed to cross-relaxation effects between NV centers of small distance to each other and NV centers very close to the diamond surface \cite{Steinert.2013, Ziem.2013}.
        We observe a similar influence of the SCO complexes on the shorter $T_1$ component.
        Examples for $T_1$ curves for the NV centers in the clean diamond and with a layer of SCO complexes applied on the diamond are shown in Fig.~\ref{fig:setup}(c).
        Due to an inhomogeneity in the laser illumination of the sample caused by the Gaussian intensity distribution of the laser beam or interference patterns, the fluorescence intensity is not homogeneously distributed over the rectangular camera sensor. 
        As a measure of the signal quality, we take the fluorescence contrast per binned pixel.
        We define the fluorescence contrast as the maximum $\Delta_{\mathrm{fl}} (\tau)$ for each binned pixel and read out the maximum contrast for each $T_1$ map. 
        In all position-resolved data sets shown below, we display regions of the $T_1$ map with a fluorescence contrast within the $1/e^2$ amplitude, while data with a contrast below the $1/e^2$ contrast is omitted.
        We provide the statistical fit errors $\Delta T_1$ for each $T_1$ map presented in the Supporting Information.
        
    	We perform widefield ODMR spectroscopy to determine the NV centers' spin transition frequency for each temperature.
    	Without an external magnetic field, the two excited $m_s$ states $m_s = \pm 1 $ of all NV centers are degenerate, respectively, and we excite all NV centers with an MW pulse at frequencies of $\SI{\approx 2866}{\giga \hertz}$.
    	To determine the duration of the MW pulse for $T_1$ measurements, we perform Rabi oscillations and choose the MW pulse duration yielding the highest fluorescence contrast. 
    	The MW pulse durations are in the order of \SI{40}{\nano \second}.
     
     \subsection{$T_2$ measurements}
     We measure the $T_2$ decoherence time with a Hahn-echo sequence \cite{Levine.2019, Barry.2020, Stanwix.2010}.
     For this, we use the same setup as depicted in Fig.~\ref{fig:setup}(a) and a sequence similar to the relaxometry sequence. 
     The pulsed sequence consists of a laser pulse of \SI{20}{\micro\second} duration, followed by MW pulses $\frac{\pi}{2}$-$\pi$-$\frac{\pi}{2}$, which are separated by waiting times $\frac{\tau}{2}$. 
     During each exposure, the sequence is repeated $N = 5000$ times.
     We accumulate 30 exposures for one picture with and without the MW pulses applied for each value of $\tau$ and repeat the entire sequence 15 times. 
     Again, we bin \qtyproduct[product-units=single]{10 x 10}{\px \squared} in the data evaluation.
     We calculate the quotient of the fluorescence signals with and without the MW pulses for each binned pixel and fit a function of type $A \exp(-\tau/T_2) + c$
     to the obtained relative fluorescence as a function of $\tau$ to extract the $T_2$ time for each binned pixel.
     In the Supporting Information, we provide the statistical fit errors $\Delta T_2$ for each $T_2$ map shown in the main text.
     We omit pixels with fit results of $R^2 < 0.95$ in the $T_2$ maps.
     We perform the Hahn echo in an external magnetic field in the order of $\SI{11}{\milli\tesla}$ generated by a permanent magnet to split the NV-center spin resonances. 
     This way, we can probe NV centers specifically with MW pulses of duration $\pi$ and $\pi/2$ in the Hahn-echo sequence at an MW frequency of \SI{\approx 2813}{\mega \hertz}.
     The duration of the $\pi$- and $\pi/2$-MW pulses are derived from Rabi oscillations.

	\section{\label{sec:Results}RESULTS AND DISCUSSION}
 	\subsection{Estimation of the effect of the paramagnetic HS Fe$^{\mathrm{II}}$ ions on the NV centers' $T_1$ time}

    \begin{figure}[h]
	\begin{overpic}[width=86mm]{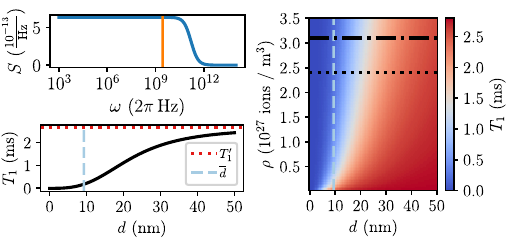}
         \put(5, 47){(a)}
         \put(49, 47){(b)}
         \put(5, 25.5){(c)}
        \end{overpic}
        \caption{\label{fig:Spectral_density}
        (a) Spectral density $S(\omega)$, calculated for Fe$^{\mathrm{II}}$ ions (blue line). The NV-center spin-transition frequency at $\omega_0 \qtyproduct[product-units=single, parse-numbers=false]{\approx 2 \pi x 2.87}{\giga \hertz}$ is marked (orange line). 
        (b) Calculated values of $T_1$ as a function of the ion density $\rho$ and the distance $d$ of the NV centers to the diamond surface. The mean distance $\bar{d}$ is marked as a blue dashed line. The value of $\rho_{\mathrm{I}} = \SI{3.1e27}{ions \per \cubic \meter}$ is marked as a black dash-dotted line, the value of $\rho_{\mathrm{II}} = \SI{2.4e27}{ions \per \cubic \meter}$ is marked as a black dotted line.
        (c) Calculated values for $T_1$ for $\rho = \rho_{\mathrm{I}}$ as a function of $d$. The $T_1$ time for the NV centers without any SCO complexes, $T_1^{\prime}$, is marked as a red dashed line.
        With $d = \bar{d}$ (blue dashed line), we expect a significant reduction of the $T_1$ time to \SI{0.18}{\milli \second}. 
       }
    \end{figure}

	Before presenting the results of the $T_1$ measurements of NV centers with the SCO complexes in proximity, we mathematically estimate the influence of a layer of paramagnetic HS Fe$^{\mathrm{II}}$ ions on the NV centers' $T_1$ time.
    The spectral density of the fluctuating magnetic field of the Fe$^{\mathrm{II}}$ ions determines their influence on the NV centers' $T_1$ time.
	It is calculated by \cite{Grant.2023, deSousa.2006}
	\begin{equation}
	    S(\omega) = \frac{2 }{\pi}\frac{\tau_c}{1 + \omega^2 \tau_c^2},
	\end{equation}
	where $\tau_c$ is the electron correlation time of Fe$^{\mathrm{II}}$.
	We calculate the ions' spectral density with $\tau_c = \SI{1e-12}{\second}$ \cite{Ducommun.1980, Mizuno.2000,  Bertini.1996, Petzold.2016} and obtain a spectral density that spreads over a wide range of frequencies, see Fig.~\ref{fig:Spectral_density}(a). 
	Since the spectral density of Fe$^{\mathrm{II}}$ overlaps with the NV-center spin-transition frequency at $\omega_0 \qtyproduct[product-units=single, parse-numbers=false]{\approx 2 \pi x 2.87}{\giga \hertz}$, we expect an influence on the NV centers' $T_1$ time due to the ions' fluctuating magnetic field. 
	The resulting $T_1$ time of the NV centers can be split into a sum of rates
	\begin{equation}
	    \frac{1}{T_1} = \frac{1}{T_1^{\prime}} + \frac{1}{T_1^{\mathrm{SCO}}},
    \label{eq:T1_SCO}
	\end{equation}
	where $T_1^{\prime}$ is the $T_1$ time of the NV centers in the diamond without any SCO complexes applied \cite{Tetienne.2013, Steinert.2013, Zhang.2023}. 
	We calculate $T_1^{\mathrm{SCO}}$ by 
	\begin{equation}
	    \frac{1}{T_1^{\mathrm{SCO}}} = 3 \gamma_e^2 B_{\perp}^2 \frac{\tau_c}{1 + \omega_0^2 \tau_c^2},
	    \label{eq:T1}
	\end{equation}
	where $B_{\perp}^2$ describes the magnetic field variance of the Fe$^{\mathrm{II}}$ ions perpendicular to the NV-center quantization axis \cite{Tetienne.2013, Steinert.2013, Zhang.2023}.
	Approximating the angle $\alpha$ between the NV-center spin and the Fe$^{\mathrm{II}}$ ions to be \SI{54.75}{\degree} in our $(100)$ surface oriented diamond (see Supporting Information), we calculate the transverse magnetic field variance by
    \begin{equation}
	    B_{\perp}^2 = \rho \left( \frac{\mu_0 \gamma_e \hbar}{4 \pi} \right)^2 \frac{S(S + 1)}{3}  (2 + 3 \sin^2\alpha) \frac{\pi}{6d^3}.
    \end{equation}
    Here, $\rho$ is the Fe$^{\mathrm{II}}$ ion density, $\mu_0$ is the vacuum permeability, $\gamma_e$ is the gyromagnetic ratio of the electron, $\hbar$ is the reduced Planck's constant, and $S$ is the total spin quantum number of the Fe$^{\mathrm{II}}$ spin.
    Since we assume the SCO layer to be directly on the diamond surface, $d$, the depth of the NV-center layer, is also the distance of the NV spin to the Fe$^{\mathrm{II}}$ layer.
    In the geometry described here, $B_{\perp}^2$ exhibits a $1/d^3$ dependence, as previously pointed out in Refs.~\cite{Steinert.2013, Ziem.2013}.
	
	Using Eq.~\ref{eq:T1}, we calculate the $T_1$ time as a function of the distance $d$ of the NV centers to the diamond surface and the Fe$^{\mathrm{II}}$ ion density $\rho$, see also Supporting Information.
	Figure~\ref{fig:Spectral_density}(b) shows the result of this calculation. 
	Starting with $T_1^{\prime} = \SI{2.8}{\milli \second}$, the $T_1$ time can drop by several orders of magnitude, depending on $d$ and $\rho$. 
	The estimated mean depth of the NV layer $\bar{d} = \SI{9.3}{\nano\meter}$ is marked as a dashed blue line in Fig.~\ref{fig:Spectral_density}(b).
	Further, we estimate $\rho$ from crystal structures of SCO complexes of the constitution or a similar constitution to \textbf{SCO~I} \cite{Grosjean.2013} and \textbf{SCO~II} \cite{Grosjean.2011}.
	From the given unit-cell volumes, we obtain $\rho_{\mathrm{I}} = \SI{3.1e27}{ions \per \cubic \meter}$ and $\rho_{\mathrm{II}} = \SI{2.4e27}{ions \per \meter \cubed}$ for \textbf{SCO~I} and \textbf{SCO~II}, respectively, which are marked in Fig.~\ref{fig:Spectral_density}(b) as horizontal lines.
	We consider these values of $\rho$ as the maximum possible ion densities since the crystal structures contain densely packed iron ions, ligands, and counter ions.
    However, in our case, we apply thin-layer samples that might exhibit a lower ion density than the crystal structures due to a different crystallization process and finite chain lengths of the polymeric structures.
	The $T_1$ time calculated as a function of $d$ for $\rho_{\mathrm{I}}$ is presented in Fig.~\ref{fig:Spectral_density}(c). 
	From our calculations, we obtain reductions of $T_1^{\prime} = \SI{2.80}{\milli \second}$ to $T_1 = \SI{0.18}{\milli\second}$ for \textbf{SCO~I} and $T_1 = \SI{0.23}{\milli\second}$ for \textbf{SCO~II} at $d = \bar{d} = \SI{9.3}{\nano \meter}$.
    This corresponds to effective magnetic-field values of $\sqrt{B_{\perp, \mathrm{I}}^2} = \SI{0.24}{\milli \tesla}$ and $\sqrt{B_{\perp, \mathrm{II}}^2} = \SI{0.21}{\milli \tesla}$ at the NV position.

	\subsection{$T_1$ measurements on SCO~I}

    \begin{figure}[h]
        \begin{overpic}[width=86mm]{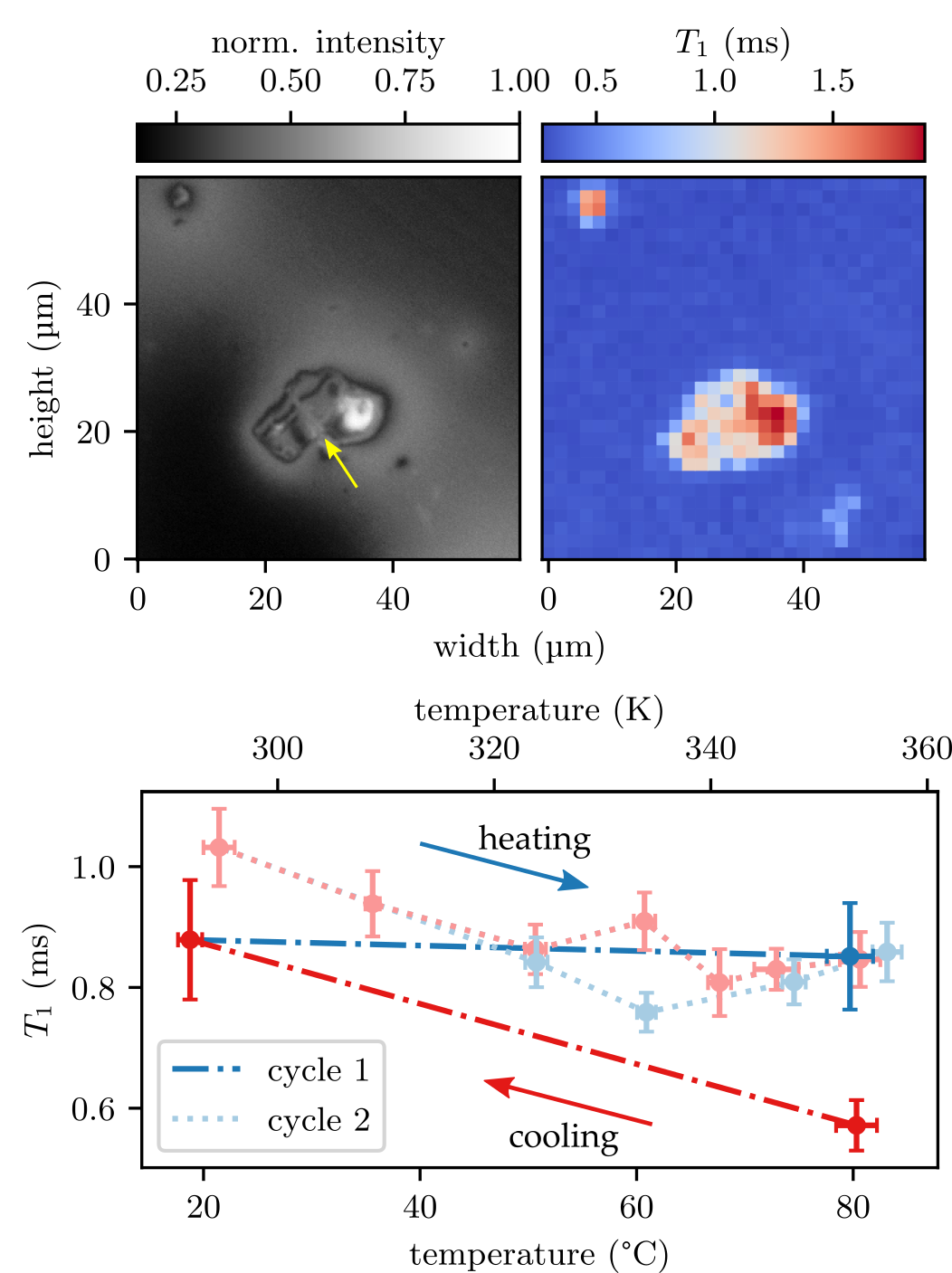}
             \put(11.5, 96){(a)}
             \put(43, 96){(b)}
             \put(5, 40){(c)}
        \end{overpic}
        \caption{\label{fig:BF4}
       $T_1$ measurements for the thin-layer sample of \textbf{SCO~I}. 
        (a) LED image of the sample \textbf{SCO~I} drop-cast on the diamond chip after several heating and cooling cycles. 
        Within the homogeneous thin layer, characteristic structures are visible. 
        The yellow arrow indicates the position investigated in (c) for different temperatures.
        (b) $T_1$ map of the same region, recorded at room temperature with widefield relaxometry. The spatial distribution of the $T_1$ times coincides with the structures in the LED image.
        (c) $T_1$ values as a function of the temperature, recorded for the position marked in the LED image with confocal microscopy for two subsequent heating and cooling cycles. 
        The red arrow indicates the cooling branch, in which the sample was heated to $\SI{115}{\degreeCelsius}$, cooled to \SI{80}{\degreeCelsius}, and measurements recorded. 
        The temperatures were then swept to room temperature, and the sample was again heated to \SI{80}{\degreeCelsius} (heating branch), and measurements were recorded.
        The errors of $T_1$ are the statistical fit errors.
        The errors of the temperatures are derived from statistical fit errors and the difference between the calculated temperature by ODMR spectroscopy and the temperature determined by the temperature sensor, see Supporting Information.
        }
    \end{figure}

    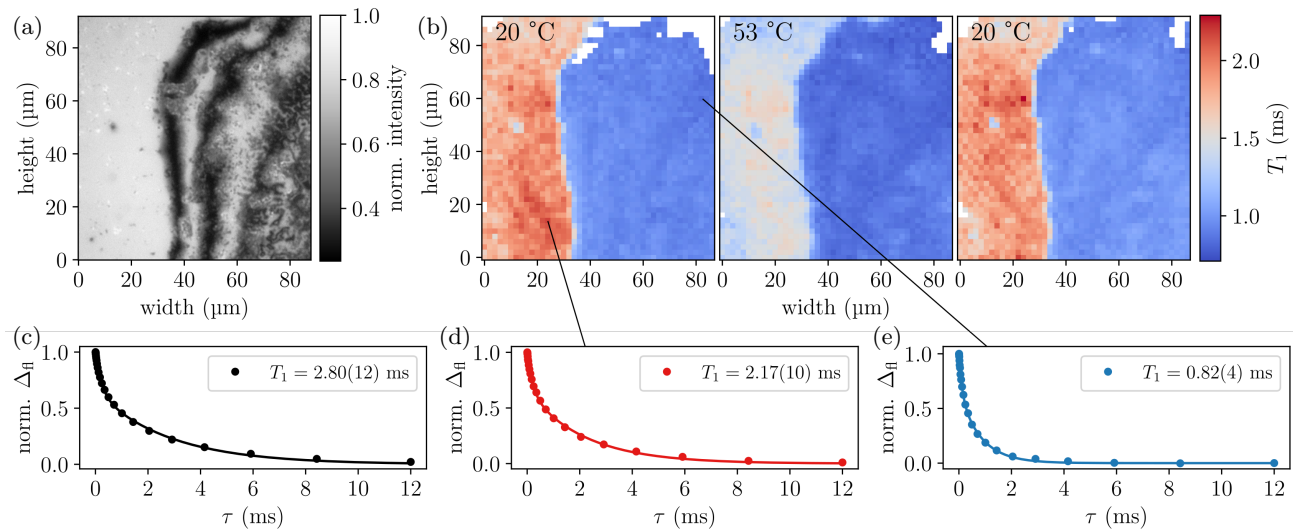
\begin{figure*}[t]
    \begin{overpic}[width=172mm]{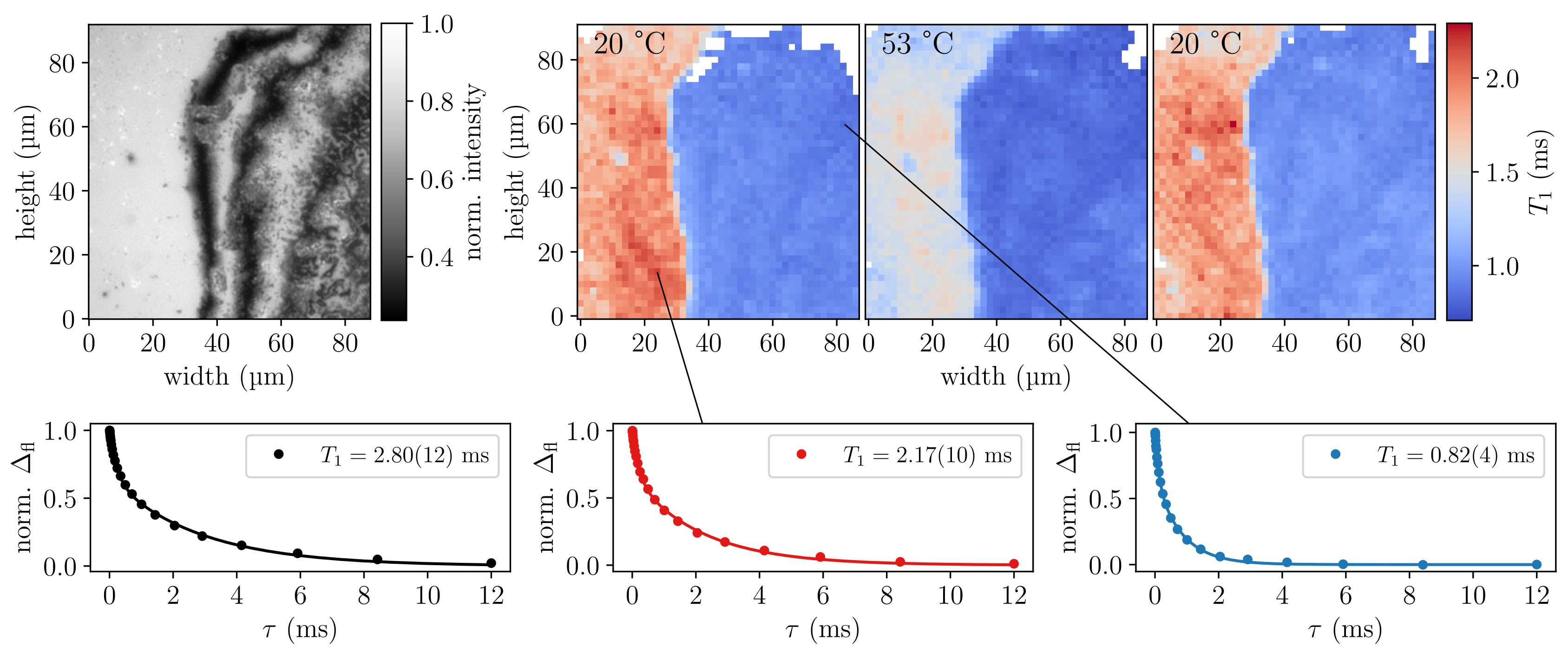}
         \put(0.5, 40){(a)}
         \put(31.8, 40){(b)}
         \put(0.5, 16){(c)}
         \put(33.5, 16){(d)}
         \put(67, 16){(e)}
        \end{overpic}
        \caption{\label{fig:MSFSO4thin}
        $T_1$ measurements for the thin-layer sample of \textbf{SCO~II}, sample 1. 
        (a) LED image of the structure visible on the thin-layer sample. 
        (b) Spatially-resolved $T_1$ measurements for the same area as shown in (a) for different temperatures. 
        The measurements were recorded in the order of appearance (left to right). 
        After cooling the sample to \SI{-20}{\degreeCelsius}, we record the $T_1$ map at \SI{20 \pm 1}{\degreeCelsius} in the heating branch. 
        Then, the $T_1$ times are measured at \SI{53 \pm 2}{\degreeCelsius}.
        Lastly, the sample is cooled back to \SI{20 \pm 1}{\degreeCelsius} to record the $T_1$ times in the cooling branch.
        Pixels of low fluorescence contrast are displayed in white (no $T_1$ time given).
        (c) $T_1$ curve (norm. $\Delta_{\mathrm{fl}}$ as a function of $\tau$) for the NV centers in the clean diamond (no SCO complexes applied) at room temperature.
        (d) $T_1$ curve for the maximum $T_1$ at \SI{20 \pm 1}{\degreeCelsius}. 
        (e) $T_1$ curve for the minimum $T_1$ at \SI{20 \pm 1}{\degreeCelsius}. 
        The solid lines in (c), (d), and (e) are fit curves as described in the methods section.
        The black lines indicate the positions in (b) at which the respective $T_1$ times are found.
        }
    \end{figure*}
    
    We apply a thin-layer sample of \textbf{SCO~I} on the diamond chip.
    Investigation of the surface under LED illumination shows visible structures in the thin layer as shown in Fig.~\ref{fig:BF4}(a), which was recorded after several cycles of heating and cooling of the sample.
    A $T_1$ measurement at room temperature of the same section is shown in Fig.~\ref{fig:BF4}(b). 
    We see an overall reduction in the $T_1$ time from $T_1^{\prime} = \SI{2.80 \pm 0.12}{\milli\second}$ to lower $T_1$ times, with a maximum of $T_1 = \SI{1.89 \pm 0.08}{\milli\second}$. 
    From the comparison of Figs.~\ref{fig:BF4}(a) and (b), it becomes clear that the visible structures in Fig.~\ref{fig:BF4}(a) correspond to regions of higher $T_1$ times, whereas sections of homogeneous appearance exhibit shorter $T_1$ times. 
    Therefore, we conclude that the structures consist of SCO material, which has lifted off the diamond chip during heating and cooling, resulting in a more considerable distance $d$ of the NV centers to the SCO sample. 
    According to our calculations, this change in $T_1$ could be caused by lifting the sample by $\SI{\approx 20}{\nano\meter}$. 
    Likewise, less SCO material at these specific positions or a combination of the two reasons might explain the longer $T_1$ times measured in these regions.
    The shortest $T_1$ time we measure in Fig.~\ref{fig:BF4}(b) is $T_1 = \SI{0.27 \pm 0.01}{\milli\second}$, which reduces $T_1^{\prime}$ by one order of magnitude. 
    This measured $T_1$ time corresponds to an effective magnetic-field value of $\sqrt{B_{\perp}^2} = \SI{0.189 \pm 0.005}{\milli \tesla}$, the uncertainty is derived from error propagation of the $T_1$-time errors. 
    The measured $T_1$ time lies in the same order of magnitude as calculated in our model, $T_1 = \SI{0.18}{\milli\second}$, which, however, assumes all Fe$^{\mathrm{II}}$ ions to be in their HS state.
    At room temperature, the complex is in its LS state and should exhibit diamagnetism \cite{Roubeau.2012, Brooker.2015}, which is not known to shorten the NV centers' $T_1$ time \cite{FreireMoschovitis.2023}.
    However, from our $T_1$ measurements, we conclude that most of the complexes are paramagnetic at room temperature.
    Our findings agree with measurements on similar SCO systems in Refs.~\cite{Lamichhane.2023, Flinn.2024}, where paramagnetism of the LS complexes was observed with ODMR spectroscopy.
    One explanation for the paramagnetism at room temperature is that terminal Fe$^{\mathrm{II}}$ ions at the ends of the polymeric chains are most likely in the HS state, independent of the temperature \cite{Wolny.2016}.
    Crystal defects between the polymeric chains additionally reduce the number of switchable Fe ions in the SCO material \cite{Dirtu.2009, Rackwitz.2012}.
    A relaxometry study on single crystals of these SCO complexes could gain insight into the chain-length dependent ratio of Fe$^{\mathrm{II}}$ ions that undergo a spin transition upon temperature variation.
    Likewise, Fe ions on the surface or on the SCO-diamond interface could favor a HS state because of partial coordination \cite{Flinn.2024, Coronado.2007}.
    Another reason for temperature-independent paramagnetism are Fe$^{\mathrm{III}}$ impurities in the SCO material, which have a similar correlation time as Fe$^{\mathrm{II}}$ \cite{Bertini.1996}, and could be caused by oxidation of the SCO material \cite{Lamichhane.2023, Flinn.2024}.
    Since the $T_1$ time is already reduced close to a minimum value at the complexes' LS state, we expect no further decrease in the $T_1$ time after heating the sample. 
    
    We probe the sample at different temperatures at a specific position as indicated with a yellow arrow in Fig.~\ref{fig:BF4}(a) in two subsequent cycles of the hysteresis and record the $T_1$ time at different temperatures, see Fig.~\ref{fig:BF4}(c).
    These experiments were performed with a confocal configuration, see Supporting Information for a detailed description.
    We first transfer the SCO sample into its HS state by externally heating it to \SI{115}{\degreeCelsius}.
    We then let the sample cool down to \SI{80}{\degreeCelsius} and record the $T_1$ time in the cooling branch (red arrow in Fig.~\ref{fig:BF4}(c)). 
    After this, we let the sample cool down to room temperature and record the $T_1$ time at the LS state of the SCO.
    To measure the $T_1$ time in the heating branch, we then heat the sample again (blue arrow in Fig.~\ref{fig:BF4}(c)).
    Raman spectroscopy suggests that the sample is in its LS state at \SI{80}{\degreeCelsius} in the heating and cooling branch. 
    It is important to note that the $T_1$ time of NV centers is known to be temperature dependent \cite{Jarmola.2012, Norambuena.2018}, see also Supporting Information. 
    Therefore, we compare the $T_1$ times pairwise in heating and cooling branches at the same temperature.
    The $T_1$ times we measure at \SI{80}{\degreeCelsius} differ significantly in the first cycle of $T_1 = \SI{0.57 \pm 0.04}{\milli \second}$ in the cooling branch to $T_1 = \SI{0.85 \pm 0.09}{\milli \second}$ in the heating branch, see Fig.~\ref{fig:BF4}(c).
    In addition to that, the change in $T_1 = \SI{0.9 \pm 0.1}{\milli\second}$ at room temperature to the shorter $T_1$ time of $T_1 = \SI{0.57 \pm 0.04}{\milli \second}$ at \SI{80}{\degreeCelsius} would be beyond the expected NV temperature effects of the $T_1$ time (\SI{\approx 0.72}{\milli \second} from Eq.~\ref{eq:T1_SCO}).
    We interpret this as a signature of a local change in the spin state of the SCO thin-layer sample.
    However, repeating our measurements, the $T_1$ times at \SI{80}{\degreeCelsius} coincide in the second cycle, while the $T_1$ times measured at \SI{60}{\degreeCelsius} differ from one another. 
    Structural changes in the SCO thin-layer sample most likely cause the changes in the $ T_1 $ times in the branches. 
    Since the SCO sample is in its LS state at \SI{80}{\degreeCelsius} in both branches, the discrepancy of the $T_1$ times in the cycles are presumably not caused by a change in the complexes' spin state.
    We observe a visible structural change in the sample during heating and cooling, see Supporting Information.
    It is known that SCO complexes change their volume when undergoing a spin transition \cite{Grosjean.2013}.
    Therefore, we assume that fractures in the SCO sample caused by repeated heating and cooling change the local environment of the NV centers and, with it, the $T_1$ time we measure.
    We conclude that structural changes in the sample and the accompanying changes in the local environment of the NV centers are prominent and dominate the measured $T_1$ times.
    Indications of a local spin change of the SCO complexes are overshadowed by this effect.

    \begin{figure*}[t]
        \begin{overpic}[width=172mm]{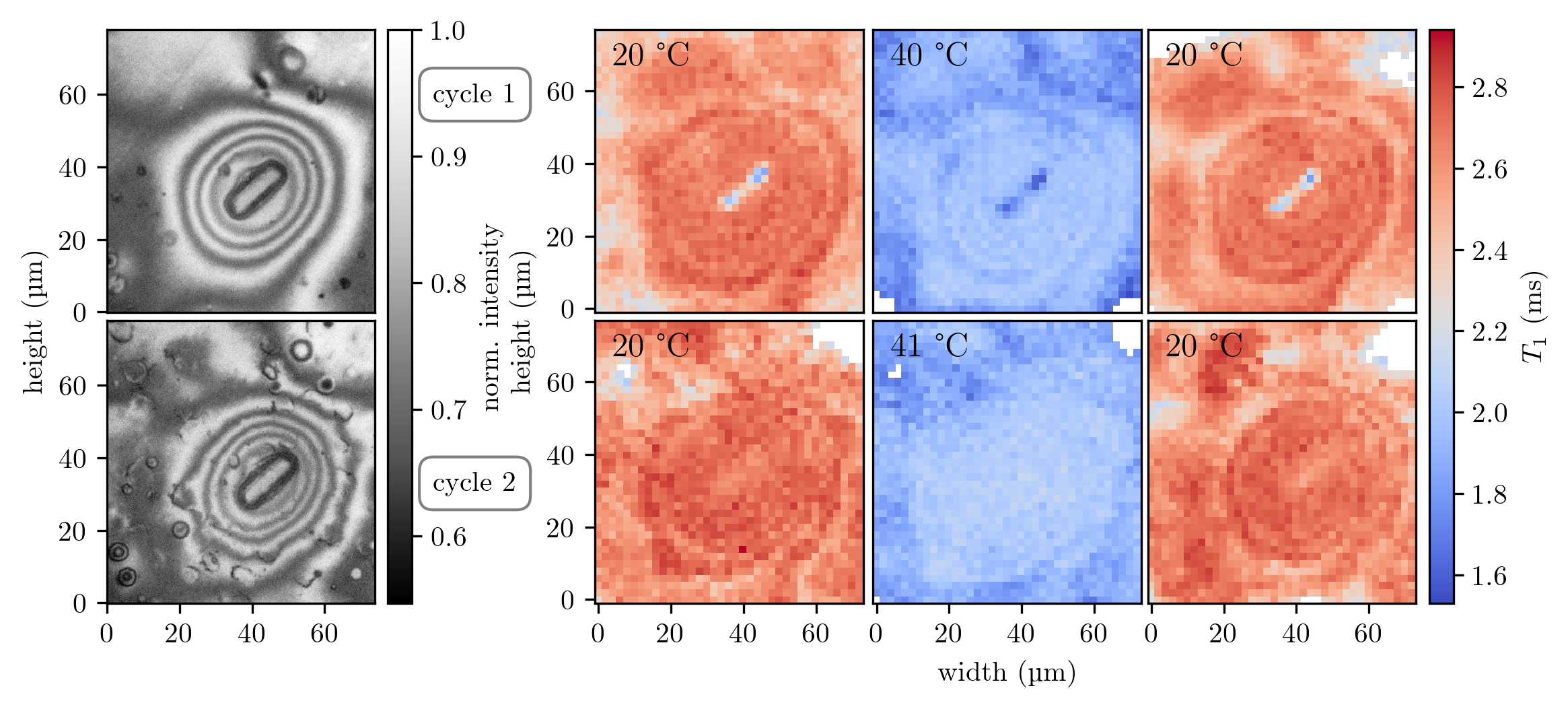}
            \put(1, 42){(a)}
            \put(32, 42){(b)}
        \end{overpic}
        \caption{\label{fig:MSFSO4thick}
        $T_1$ measurements for the thin-layer sample 2 of \textbf{SCO~II} for two subsequent heating and cooling cycles. The first row (second row) displays the data for the first cycle (second cycle).
        (a) LED image of the structure visible on the thin-layer sample at the beginning of the first and second cycles.
        Comparing the two images, heating and cooling the sample has caused cracks in the sample from the first to the second cycle.
        (b) Spatially-resolved $T_1$ measurements for the same area as shown in (a) for different temperatures for the first and second cycles. 
        The measurements were recorded in the order of appearance (left to right). 
        After cooling the sample to \SI{-20}{\degreeCelsius}, we record the $T_1$ map at \SI{20 \pm 1}{\degreeCelsius} in the heating branch. 
        Then, the $T_1$ times are measured at \SI{40 \pm 2}{\degreeCelsius} (\SI{41 \pm 2}{\degreeCelsius}) for cycle 1 (cycle 2).
        Lastly, the sample is cooled back to \SI{20 \pm 1}{\degreeCelsius} to record the $T_1$ times in the cooling branch.
        While the rectangular structure in (a) exhibits a lower $T_1$ time than its surroundings in the first cycle, this $T_1$ contrast fades in the second cycle.
        Pixels of low fluorescence contrast are displayed in white (no $T_1$ time given).
        }
    \end{figure*}

    \subsection{$T_1$ measurements on SCO~II}

    Since the spin-switching temperature for \textbf{SCO~I} lies beyond the temperatures at which we can record measurements, we apply another SCO compound to our NV diamond, \textbf{SCO~II}, which switches its spin state at significantly lower temperatures. 
    From Raman spectroscopy, we find that the complex is \SI{62}{\percent} in its HS state at room temperature.
    In contrast, above \SI{35}{\degreeCelsius}, the complexes switch to \SI{84}{\percent} HS, where they remain after cooling to room temperature, see Supporting Information.
    According to our model, this change in the density of Fe$^{\mathrm{II}}$ ions in the HS state should be accompanied by a significant decrease in the $T_1$ time, see below.
    We investigate two samples of \textbf{SCO~II}, which differ by the amount of SCO material applied, see Sec.~\ref{sec:methods}.

    In the first sample, a comparable amount of \textbf{SCO~II} is applied on the NV-diamond chip as we presented previously for \textbf{SCO~I}.
    An LED image and $T_1$ data for different temperatures can be seen in Fig.~\ref{fig:MSFSO4thin}(a).
    In the image, a dark shade is visible on the right half of the image. 
    We interpret this as a higher density of SCO material on the diamond sample.
    Recording the $T_1$ times in this specific region, we find that the $T_1$ time is reduced compared to $T_1^{\prime}$ in the entire area, see Fig.~\ref{fig:MSFSO4thin}(b) in comparison to Fig.~\ref{fig:MSFSO4thin}(c). 
    We find a maximum of $T_1 = \SI{2.17 \pm 0.10}{\milli\second}$, see Fig.~\ref{fig:MSFSO4thin}(d), and a minimum of $T_1 = \SI{0.82 \pm 0.04}{\milli\second}$ at \SI{20}{\degreeCelsius}, see Fig.~\ref{fig:MSFSO4thin}(e), indicating a paramagnetic sample.
    We also observe that the $T_1$ time is especially short in the right half of the observed area, while it is up to an order of magnitude longer on the left half. 
    This coincides with the intensity distribution in the LED image and underlines our notion that a higher density of SCO material is present in the dark-shaded areas of the image, leading to a reduced $T_1$ time. 
    We heat the complexes beyond the spin transition temperature and perform $T_1$ measurements again at \SI{53}{\degreeCelsius}. 
    The contrast in $T_1$ times between the left and right side of the map persists, however, it is diminished due to temperature effects in the diamond, influencing the $T_1$ time  of the NV centers, see Fig.~\ref{fig:MSFSO4thin}(b). 
    We find a reduction to a maximum of $T_1 = \SI{1.69 \pm 0.07}{\milli \second}$ and a minimum of $T_1 = \SI{0.71 \pm 0.04}{\milli \second}$, which is in the order of what we expect from the temperature dependence of the $T_1$ time for the bare diamond, see Supporting Information.
    To separate temperature effects and effects due to the spin switching of the SCO material, we cool the sample back to \SI{20}{\degreeCelsius} and repeat the $T_1$ measurements in the cooling branch. 
    We find that the $T_1$ times coincide with the $T_1$ times measured previously in the heating branch at \SI{20}{\degreeCelsius}, with a maximum of $T_1 = \SI{2.29 \pm 0.10}{\milli\second}$ and a minimum of $T_1 = \SI{0.82 \pm 0.04}{\milli\second}$. 
    Calculated from our model, a change of the HS-Fe density from \SI{62}{\percent} to \SI{84}{\percent} should have reduced the $T_1$ minimum to $T_1 = \SI{0.66}{\milli \second}$, which we do not observe. 
    Temperature-independent paramagnetism of the SCO samples prevail and overshadow the effects on the $T_1$ time of the spin transition.
    We observe a similar behavior in a second cycle, see Supporting Information.

     \begin{figure*}[t]
        \begin{overpic}[width=172mm]{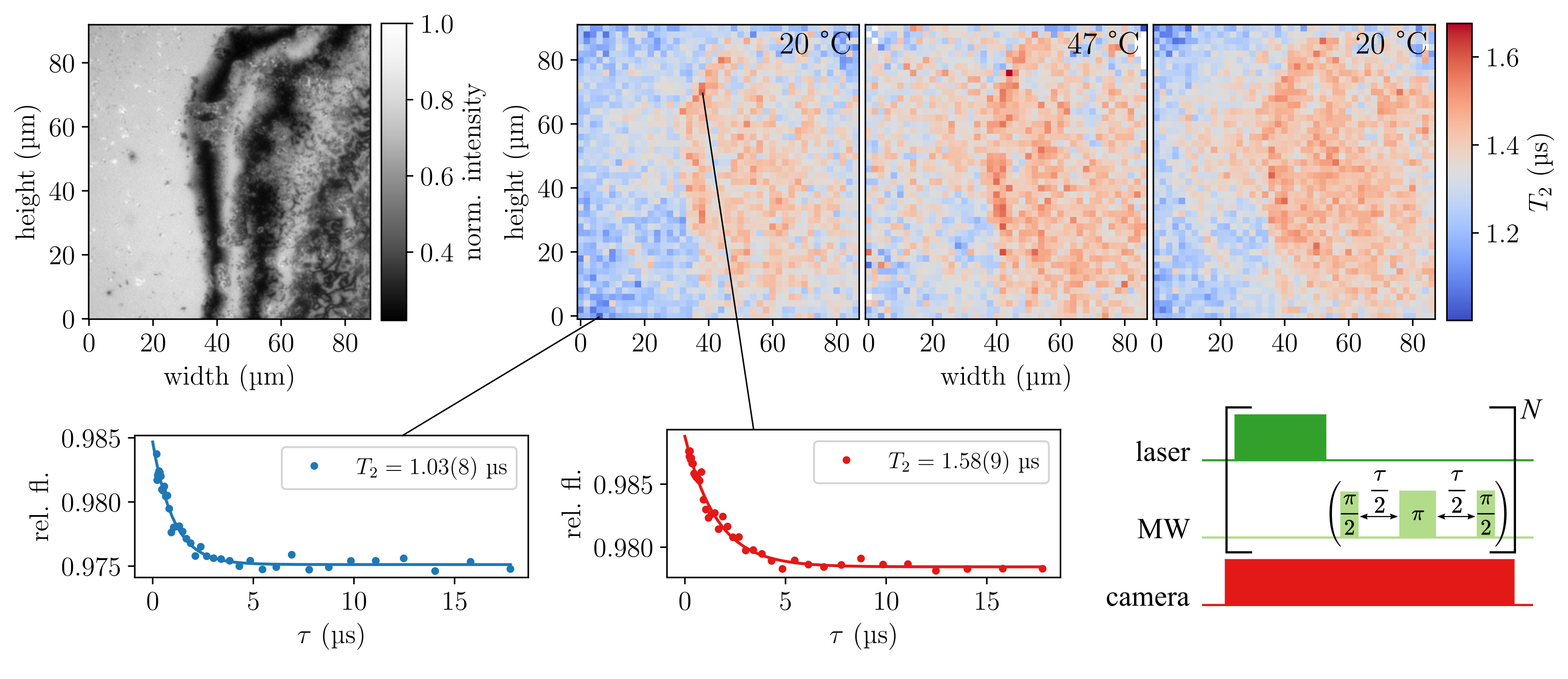}
            \put(0.5, 42){(a)}
            \put(31.8, 42){(b)}
            \put(1, 16){(c)}
            \put(35, 16){(d)}
        \end{overpic}
        \caption{\label{fig:MSFSO4Hahn}
       $T_2$ measurement (Hahn echo) for the thin-layer sample 1 of \textbf{SCO~II} in the third cooling and heating cycle. 
        We use a measurement sequence as shown and described in the methods section.
        (a) LED image of the structure examined. 
        (b) Spatially-resolved $T_2$ measurements for the same area as shown in (a) for different temperatures. 
        The data was recorded in order of appearance (left to right). 
        After cooling the sample to \SI{-20}{\degreeCelsius}, we record the $T_2$ times at \SI{20 \pm 1}{\degreeCelsius} in the heating branch. Next, we heat the sample to \SI{47 \pm 4}{\degreeCelsius}, repeat the measurement, and let it cool to \SI{20 \pm 1}{\degreeCelsius} to repeat the measurement in the cooling branch.
        (c) Relative fluorescence (rel.\,fl.) as a function of $\tau$ for the minimum $T_2$ at \SI{20 \pm 1}{\degreeCelsius}.
        (d) Relative fluorescence as a function of $\tau$ for the maximum $T_2$ at \SI{20 \pm 1}{\degreeCelsius}.
        The solid lines in (c) and (d) are fit curves as described in the methods section. 
        The black lines indicate the positions at which the respective $T_2$ times are found.
        }
    \end{figure*}

    To test the effect of a thicker SCO layer on the NV centers' $T_1$ time, we repeat the measurements on a second sample of \textbf{SCO~II}.
    Here, we apply approximately twice the amount of SCO material on the diamond chip compared to sample 1. 
    The results for this sample can be seen in Fig.~\ref{fig:MSFSO4thick} for two subsequent cycles.
    Again, we choose a particular spot of the sample with visible characteristics in the thin layer. 
    The structure investigated is rectangular, surrounded by a circular pattern, see Fig.~\ref{fig:MSFSO4thick}(a).
    We find that the $T_1$ map at \SI{20}{\degreeCelsius} again shows similar spatial characteristics as the LED image of the area.
    At room temperature, we find that the $T_1$ time in the observed area is approximately the same as $T_1^{\prime}$, with a maximum of $T_1 = \SI{2.86 \pm 0.12}{\milli\second}$, see Fig.~\ref{fig:MSFSO4thick}(b). 
    In contrast, the rectangular structure exhibits a notably shorter $T_1$ time than its surroundings, with a minimum of $T_1 = \SI{1.86 \pm 0.09}{\milli\second}$. 
    We heat the sample above the SCO spin-transition temperature to \SI{40}{\degreeCelsius} and repeat the $T_1$ measurements, see the first row of Fig.~\ref{fig:MSFSO4thick}(b). 
    The $T_1$ time decreases overall, showing a maximum of $T_1 = \SI{2.08 \pm 0.09}{\milli \second}$. 
    The minimum $T_1$ time within the rectangular structure lies at $T_1 = \SI{1.59 \pm 0.06}{\milli\second}$. 
    The decrease in the $T_1$ time in this area is in the order of $T_1$ reduction expected due to temperature effects of the bare diamond.
    Separating $T_1$ temperature effects from possible SCO effects, we let the sample cool down to room temperature and repeat the measurement.
    We find that the $T_1$ time is restored to its values in the heating branch with a maximum of $T_1 = \SI{2.82 \pm 0.11}{\milli\second}$ and a minimum of $T_1 = \SI{1.81 \pm 0.08}{\milli\second}$, which lies within the rectangular structure.
    We probe the sample in a second cycle, see the second row of Fig.~\ref{fig:MSFSO4thick}. 
    After cooling the sample, the structure has visibly changed, see Fig.~\ref{fig:MSFSO4thick}(a). 
    Cracks have appeared, especially in the surroundings of the rectangular structure. 
    We record the $T_1$ map for this area and find that the rectangular structure no longer exhibits a shorter $T_1$ time than its surroundings and find a mean $T_1 = \SI{2.64 \pm 0.12}{\milli \second}$ for the entire area at room temperature, which coincides with $T_1^{\prime}$. 
    At increased temperature, we obtain a mean $T_1 = \SI{1.95 \pm 0.09}{\milli \second}$.
    Cooling back to room temperature, we measure $T_1 = \SI{2.62 \pm 0.13}{\milli \second}$, where the errors denote the standard deviations.
    We assume that the SCO complexes of rectangular shape lifted off the diamond chip after the second cooling process, causing the $T_1$ time to increase in the second cycle. 
    We observe no effect on the NV centers' $T_1$ time due to the spin switching of the SCO complexes.
    We note that in a second position on the same sample, we observe a similar behavior, see Supporting Information.
    These results underline that the method applied here is susceptible to small changes in the distance of the SCO complexes to the diamond surface. 
    One reason for the overall higher $T_1$ time in this sample compared to the previous ones might be that the thicker SCO layer has poor contact with the diamond chip and tends to detach from the substrate when cooling and heating, which results in a larger mean distance of the SCO layer to the NV layer.
    We observe that even thicker layers of the complexes chip off the diamond after heating or cooling completely, making $T_1$ measurements with thicker SCO layers inaccessible.

    While the presented measurements show a spatial distribution of $T_1$ times that coincides with the visible structure of the SCO material, no spin switching of the SCO complexes can be demonstrated with our method.
    Compared to \textbf{SCO~I}, which is in its LS state at the temperatures we probed, \textbf{SCO~II} is mostly in its HS state. 
    Nevertheless, the $T_1$ times we observe for \textbf{SCO~II} are still in the same order of magnitude as for \textbf{SCO~I}, although both complexes should exhibit different amounts of HS Fe$^{\mathrm{II}}$ ions.
    We conclude from this that paramagnetic centers within the SCO complexes in their LS state already dominate the result of the $T_1$ measurements, causing the method to be unable to detect additional changes in the spin state of the SCO complexes.

    \subsection{$T_2$ measurements on SCO~II}
    
    While $T_1$ relaxometry is known to detect fluctuating magnetic fields in the GHz range, Hahn echoes are sensitive towards lower-frequency fluctuating magnetic fields in the MHz range \cite{SchaferNolte.2014}. 
    To shift the detection window to these lower frequencies, we perform a Hahn echo on \textbf{SCO~II}, sample 1, at the same position as previously shown for $T_1$ relaxometry in Fig.~\ref{fig:MSFSO4thin} in a third cycle.
    We perform these experiments in an external magnetic field ($\SI{\approx 11}{\milli\tesla}$) to precisely control NV centers of one orientation in the diamond lattice with MW pulses and probe the NV centers with MW frequencies of $\SI{\approx 2813}{\mega\hertz}$.
    Since the spectral density of the magnetic field of the Fe$^{\mathrm{II}}$ ions is distributed over a wide frequency range, see Fig.~\ref{fig:Spectral_density}(a), the Hahn echo should be able to detect the fluctuating magnetic field arising from the paramagnetic SCO complexes. 
    We show the measurement sequence and the results for different temperatures in Fig.~\ref{fig:MSFSO4Hahn}.
    The LED image of the structure in Fig.~\ref{fig:MSFSO4Hahn}(a) shows the same characteristics as previously presented in Fig.~\ref{fig:MSFSO4thin}. 
    We note that the measured $T_2$ times are three orders of magnitude lower than the $T_1$ times we measure, see Fig.~\ref{fig:MSFSO4Hahn}(b). 
    The spatial distribution of the $T_2$ times coincides with the structures we see in the LED image in Fig.~\ref{fig:MSFSO4Hahn}(a).
    However, the areas of the different densities of SCO material are not as clearly separated by differences in $T_2$ as in the $T_1$ measurements.
    We conclude from this that the Hahn echo is less sensitive towards the fluctuating magnetic field of the paramagnets, which conforms with the observations made by Ref.~\cite{Steinert.2013}.
    In addition, we do not observe any changes in the $T_2$ times upon temperature variation.
    While the $T_1$ measurements in Fig.~\ref{fig:MSFSO4thin} show a reduced $T_1$ time in darker areas of the sample in comparison to brighter areas, we find that the opposite is the case for the $T_2$ time, see Fig.~\ref{fig:MSFSO4Hahn}(b).
    We find a minimum of $T_2 = \SI{1.03 \pm 0.08}{\micro\second}$ and a maximum of $T_2 = \SI{1.58 \pm 0.09}{\micro\second}$, see Figs.~\ref{fig:MSFSO4Hahn}(c) and (d), at the indicated positions.
    In addition to that, we observe that the $T_2$ time in the right half of the area is slightly increased in comparison to the $T_2$ time of the NV centers without SCO complexes of $T_2^{\prime} = \SI{1.26 \pm 0.05}{\micro \second}$, see Supporting Information.
    These contrasting results between $T_1$ and $T_2$ measurements might be caused by changes in the detection sensitivity of the NV center for Fe$^{\mathrm{II}}$ or Fe$^{\mathrm{III}}$ in the significantly different detection windows of the measurement sequences. 
    Further investigations on this topic are necessary, which could involve tuning the $T_1$ detection window by application of an external magnetic field or implementation of more complex pulse sequences, such as CPMG (Carr-Purcell-Meiboom-Gill) sequences, which are known to show higher sensitivities towards the detection of paramagnetic noise than the Hahn echo \cite{Steinert.2013}.
    In addition, the systematic influence of nanoscale distance changes between SCO layer and NV layer on $T_1$ and $T_2$ measurements can be studied to gain further insight into the fluctuating magnetic fields of Fe$^{\mathrm{II}}$ and Fe$^{\mathrm{III}}$. 

	\section{\label{sec:Concl}CONCLUSIONS}
    In this work, we probe the fluctuating magnetic field of SCO thin-layer samples with NV centers in diamond implanted at \SI{9.3}{\nano \meter} depth.
    For this, we perform spatially resolved NV-center relaxometry to quantify influences of the paramagnetic Fe ions on the $T_1$ times at different temperatures.
    From our results, we conclude that paramagnetic centers dominate the results of the $T_1$ measurement for the LS complexes.
    We initially observe indications of a local change in the SCOs' spin state.
    However, the temperature variation causes structural changes in the thin-layer sample, affecting the $T_1$ time due to an altered sample-NV distance.
    These effects are so prominent that an additional change in the SCO complexes' spin state from LS to HS cannot be detected.
    These findings underline the high sensitivity of NV-center relaxometry for the detection of paramagnetic materials and the high susceptibility towards changes in the distance of the sample to the NV layer.

    \begin{acknowledgments}
        We acknowledge support from the Nano Structuring Center (NSC) of the RPTU Kaiserslautern-Landau.
        This project was funded by the Deutsche Forschungsgemeinschaft
        (DFG, German Research Foundation)—Project-ID No. 454931666 in the initial stages and QUIP (Quanten-Initiative Rheinland-Pfalz). 
        Further, I.~C.~B. thanks the Studienstiftung des deutschen Volkes.
        T.~H., J.~A.~W., and V.~S. acknowledge the support by the Deutsche Forschungsgemeinschaft (DFG) through CRC/ TRR173, “Spin + X”, project A4.
        We thank O. Opaluch and E. Neu-Ruffing (RPTU Kaiserslautern), M. Schreck and W. Brückner (Augsburg University) for diamond ion implantation. 
        Further, we thank O. Opaluch for the diamond sample preparation. 
        We thank M. Weiler, Z. Zhang,  N. Mathes, and F. Freire-Moschovitis for helpful discussions, and J. Witzenrath for experimental support.
        We would like to express special gratitude to E. Neu-Ruffing and O. Opaluch, who have given us valuable experimental support during our work. 
    \end{acknowledgments}

    \FloatBarrier

    %
    
    \clearpage

    \section*{SUPPORTING INFORMATION}
    \setcounter{equation}{0}
    \setcounter{section}{0}
    \setcounter{figure}{0}
    \renewcommand{\theequation}{S\arabic{equation}}
    \renewcommand{\thetable}{S\arabic{table}}
    \renewcommand{\thesection}{\arabic{section}}
    \renewcommand{\thefigure}{S\arabic{figure}}
    \renewcommand{\bibnumfmt}[1]{[S#1]}
    \renewcommand{\citenumfont}[1]{S#1}
    
    \subsection{Estimation of $T_1$}
    In our experiments, SCO complexes are brought to proximity with NV centers in diamond.
    The SCO complexes consist of Fe$^{\mathrm{II}}$ ions, which are paramagnetic when in their high-spin (HS) state. 
    This paramagnetic spin bath at the NV center location is characterized by a zero-mean fluctuating magnetic field $\mathbf{B}(t)$, which will influence the NV centers' longitudinal spin relaxation time $T_1$. 
    More precisely, the NV centers' longitudinal spin relaxation is influenced by transverse magnetic-field components of the paramagnetic ions with spectral-density amplitude at the NV center transition frequency \cite{Tetienne.2013_S}.
    The dynamics of the fluctuating magnetic field caused by molecules or ions with an unpaired electron spin can be described with the correlation function as \cite{deSousa.2006_S, Steinert.2013_S}
    \begin{equation}
	       \langle B_k(0)B_k(\tau)\rangle = \langle B_k^2 \rangle \exp(-|\tau|/\tau_c).
    \end{equation}
    Here, $\langle B_k^2 \rangle$ ($k = x, y, z$) is the variance of the magnetic field, and the correlation time $\tau_c$ of the magnetic field describes the memory of the noise generated by the paramagnetic ions \cite{deSousa.2006_S}.
    The spectral noise density of the magnetic field is given by the Fourier transform of its correlation function, and summarizes the properties of the fluctuating magnetic field. 
    The normalized spectral density is of form \cite{Grant.2023_S, deSousa.2006_S}
    \begin{equation}
	   S(\omega) = \frac{2}{\pi} 
	   \frac{\tau_c}{1 + \omega^2 \tau_c^2},
    \end{equation}
    which has the shape of a Lorentzian.
    In the literature, one finds that $\tau_c$ for Fe$^{\mathrm{II}}$ lies in the order of \SI{1e-12}{\second} \cite{Ducommun.1980_S, Mizuno.2000_S,  Bertini.1996_S, Petzold.2016_S}. 
    From this, we calculate the spectral density of the magnetic field generated by the HS Fe$^{\mathrm{II}}$ ions.
    The calculated normalized spectral noise density for Fe$^{\mathrm{II}}$ is shown in the main text in Fig.~\ref{fig:Spectral_density}(a).
    Depending on the quantum sensing scheme, the NV center can detect magnetic noise from sub-Hz over kHz, to MHz, and GHz frequencies \cite{SchaferNolte.2014_S}. 
    While, for example, Hahn echoes can be applied to sense noise in the $\SI{}{\mega \hertz}$ range, $T_1$ sensing is used to detect noise in the $\SI{}{\giga \hertz}$ range \cite{SchaferNolte.2014_S, Steinert.2013}. 
    To quantify this, the overlap of the ions' spectral density $S(\omega)$ and the NV-center filter function for $T_1$ relaxometry $F(\omega)$ is considered of form $\int S(\omega)F(\omega)\,\mathrm{d}\omega$ \cite{SchaferNolte.2014_S, Grant.2023_S}. 
    The filter function $F(\omega)$ for $T_1$ relaxometry is given by
    \begin{equation}
    	F(\omega) = \frac{1}{\pi} \frac{\Gamma}{\Gamma^2 + (\omega - \omega_0)^2},
    \end{equation}
    where $\Gamma = 1/T_2^*$ is the NV spin dephasing rate and $\omega_0$ is the NV transition frequency \cite{SchaferNolte.2014_S}.
    With the spectral density of the magnetic noise generated by the paramagnetic ions overlapping with the NV centers' spin transition frequency, the NV centers' $T_1$ time will be reduced due to the ion's fluctuating magnetic field.
    The NV center's relaxation rate $1/T_1$ under this influence is described by the relaxation rate $1/T_1^\prime$ of the diamond with intrinsic impurities and an additional rate $1/T_1^{\mathrm{SCO}}$ proportional to  spectral density as \cite{Tetienne.2013_S, Zhang.2023_S}  
    
    \begin{equation}
    	\frac{1}{T_1} = \frac{1}{T_1^{\prime}} + \frac{1}{T_1^{\mathrm{SCO}}}.
    \end{equation}
    Because $1/\tau_c = 1/(\SI{1e-12}{\second}) \gg \Gamma \approx 1/(\SI{50e-9}{\second})$, the rate for spin relaxation can be approximated by
    
    \begin{equation}
    	\frac{1}{T_1} = \frac{1}{T_1^{\prime}} + A \frac{\gamma_e^2}{2} \left[S_{B_x} ( \omega_0) + S_{B_y} (\omega_0) \right], 
    \end{equation}
    taking into account the transverse components of the magnetic field.
    The spectral densities of the magnetic-field components are given by  \cite{Zhang.2023_S, Tetienne.2013_S}
    
    \begin{equation}
    	S_{B_{x, y}} ( \omega_0) = \int_{-\infty}^{\infty} B_{x, y} (t) B_{x, y}(t+ \tau) \exp(-i \omega_0 \tau)\,\mathrm{d}\tau.
    \end{equation}
    Here, $A = 3$ is related to the rate equations in an $S = 1$ system \cite{Tetienne.2013_S}, $\gamma_e \approx 2 \mu_\mathrm{B} / \hbar$ is the gyromagnetic ratio of the electron, $\mu_\mathrm{B}$ is Bohr's magneton, and $\hbar$ is the reduced Planck constant.
    Further, $\omega_0 = 2\pi \times \SI{2.87}{\giga\hertz}$, is the NV center's spin transition frequency.
    The integration yields an expression of the transverse magnetic-field variance as \cite{Tetienne.2013_S, Zhang.2023_S}
    
    \begin{align}
    	\frac{1}{T_1} &= 
    	\frac{1}{T_1^{\prime}} + A \frac{\gamma_e^2}{2} \left( \langle B_x^2 \rangle + \langle B_y^2 \rangle \right) \frac{2\tau_c}{1 + \omega_0^2 \tau_c^2} \nonumber \\ &= 
    	\frac{1}{T_1^{\prime}} + A \gamma_e^2 B_{\perp}^2 \frac{\tau_c}{1 + \omega_0^2 \tau_c^2}.
    \end{align}
    Here, $B_\perp^2$ is the variance of the magnetic field perpendicular to the NV quantization axis ($z$ axis).
    The transverse magnetic-field variance is given by the sum of the magnetic-field variance contributions of each spin $i$ \cite{Tetienne.2013_S}
    \begin{equation}
    	B_{\perp}^2 = \sum_i B_{\perp, i}^2,
    \end{equation}
    
    while each magnetic-field variance can be calculated by \cite{Tetienne.2013_S, Zhang.2023_S}
    
    \begin{equation}
    	B_{\perp, i}^2 = \left( \frac{\mu_0 \gamma_e \hbar}{4 \pi} \right)^2 \frac{S(S + 1)}{3} \frac{(2 + 3 \sin^2\alpha_i)}{r_i^6}. 
    \end{equation}
    
\begin{figure}[h]
    	\includegraphics[width=86mm]{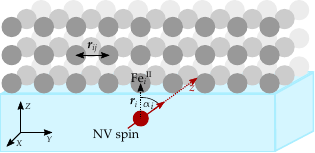}
            \caption{\label{fig:calculation_Fe_T1}
          	Schematic drawing of the SCO complexes on the diamond chip. The NV center spin lies at \SI{54.75}{\degree} to the diamond surface.
    		The vector to a nearby Fe$^{\mathrm{II}}$ spin is termed $\mathbf{r}_i$, and the angle between $\mathbf{r}_i$ and the NV axis is termed $\alpha_i$. 
    		Two Fe$^{\mathrm{II}}$ centers are separated by $\mathbf{r}_{ij}$.
            The figure was drawn after the description in Ref.~\cite{Tetienne.2013_S}.
            The NV-quantization axis is marked as the $z$ axis, while the laboratory coordinate system is specified with capital letters.
           }
        \end{figure}

    In high spin, the spin quantum number $S$ is $S = 4/2$ for Fe$^{\mathrm{II}}$ ions.
    Further, $\mu_0$ is the vacuum permeability.  
    The angle $\alpha_i$ is the angle between $\mathbf{r}_i$, which points from the NV center to an Fe$^{\mathrm{II}}$ spin, and the NV's quantization axis, see Fig.~\ref{fig:calculation_Fe_T1}.
    To estimate the transverse magnetic field components, we microscopically model the influence of the Fe$^{\mathrm{II}}$ ions.
    The Fe$^{\mathrm{II}}$ ions in the polymeric SCO complexes are ordered chain-like. 
    To estimate the volume density of Fe$^{\mathrm{II}}$ ions in the SCO samples, we extract the unit-cell volumes of the crystal structure for \textbf{SCO~I} \cite{Grosjean.2013_S} and of a crystal structure similar to \textbf{SCO~II} \cite{Grosjean.2011_S}.
    In the case of \textbf{SCO~I}, a unit cell of volume $V = \SI{1303.2}{\angstrom \cubed}$ contains four HS Fe$^{\mathrm{II}}$ ions.
    We calculate the volume density to $\rho_{\mathrm{I}} = \SI{3.1e27}{ions\,\per \meter \cubed}$.
    A unit cell of the structure similar to \textbf{SCO~II} contains two Fe$^{\mathrm{II}}$ ions and is of volume $V = \SI{834.3}{\angstrom \cubed}$. 
    Therefore, a volume density $\rho_{\mathrm{II}}$ of Fe$^{\mathrm{II}}$ ions is given by $\rho_{\mathrm{II}} = \SI{2.4e27}{ions\,\per \meter \cubed}$. 
    Assuming a mean angle $\alpha_i(X, Y, Z) \approx \alpha$, we write the transverse magnetic-field variance as
    \begin{multline}
    	B_{\perp}^2 \approx \rho \left( \frac{\mu_0 \gamma_e \hbar}{4 \pi} \right)^2 \frac{S(S + 1)}{3}  (2 + 3 \sin^2\alpha) \\
        \int_{d}^{\infty} \int_{-\infty}^{\infty} \int_{-\infty}^{\infty}  \frac{1}{r^6}\,\mathrm{d}X\mathrm{d}Y\mathrm{d}Z,
    \end{multline}
    with $r = \sqrt{X^2 + Y^2 + Z^2}$ and $d$ the depth of the NV layer in the diamond, yielding
    \begin{equation}
    	B_{\perp}^2 = \rho \left( \frac{\mu_0 \gamma_e \hbar}{4 \pi} \right)^2 \frac{S(S + 1)}{3}  (2 + 3 \sin^2\alpha) \frac{\pi}{6d^3}.
    \end{equation}
    Since our experiments are performed on a diamond with face orientation $(100)$, we assume $\alpha = \SI{54.75}{\degree}$ in our calculations.
    We have verified in numerical simulations that this fixed angle yields similar results to summing all individual contributions with varying distance and angle from an array of Fe$^{\mathrm{II}}$ ions as shown in Fig.~\ref{fig:calculation_Fe_T1}. 
    Importantly, this approximation is only valid for the $(100)$ diamond, as we find discrepancies between the approximate analytical solution and a numerical solution for an NV spin oriented perpendicular to the surface.
    Using a simulation (``The Stopping and Range of Ions in Matter'' (SRIM) \cite{SRIM_S}), we determine the mean depth of the NV centers in the diamond to $\bar{d} = \SI{9.3}{\nano\meter}$.
    
    It follows that for a distance of the NV centers from the diamond surface of $d = \SI{9.3}{\nano\meter}$, the $T_1$ time of the NV centers will be reduced from $T_1^{\prime} = \SI{2.80}{\milli\second}$ to $T_1 = \SI{0.18}{\milli\second}$ with HS Fe$^{\mathrm{II}}$ ions of \textbf{SCO~I} and to $T_1 = \SI{0.23}{\milli\second}$ of \textbf{SCO~II }deposited on the diamond, which are reductions of the relaxation time by one order of magnitude.
    
    \subsection{Confocal setup}
\begin{figure}[h]
    	\includegraphics[width=86mm]{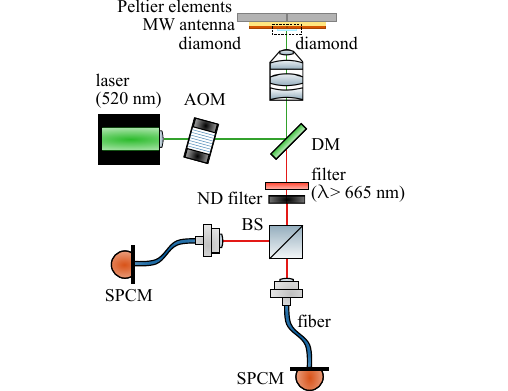}
            \caption{\label{fig:confocal}
          	Confocal configuration of the microscope setup. 
            Compared to the widefield setup, a collimated laser beam enters the objective and is then focused onto the sample.
            The NV fluorescence $\SI{>665}{\nano \meter}$ is detected with single-photon counting modules (SPCMs).
            Neutral-density (ND) filters are used to keep the SPCMs below saturation.
           }
        \end{figure}

    Our confocal setup is similar to our widefield setup, see Fig.~\ref{fig:confocal}.
    We use a 520-nm laser of power of $\SI{\approx 800}{\micro \watt}$.
    In the laser beam path, we remove the lens in front of the objective, so that the collimated laser beam enters its aperture.
    The laser beam is then focused onto the NV-diamond sample.
    We estimate the laser focus to be $\SI{\approx 1}{\micro \meter}$ wide (1/$e^2$ diameter).
    The NV fluorescence is filtered, and wavelengths $\SI{>665}{\nano \meter}$ are detected with two single-photon counting modules (SPCMs).
    To keep the SPCMs below saturation, we employ neutral-density (ND) filters. 
    For $T_1$ measurements, we use a sequence that is similar to the one described in our previous work \cite{CardosoBarbosa.2023_S}.
    Combined with a time-to-digital converter, we time-resolve the fluorescence during the relaxometry sequence and calculate a $T_1$ curve, as mentioned in the main text, by forming the mean signal of both SPCMs. 
    We fit the data as described in the main text.
    
    \subsection{Temperature determination}
    We heat our sample with Peltier elements and increase the temperature by setting a current. 
    On the surface of the holder, a temperature sensor is mounted. 
    Via ODMR spectroscopy, we measured the zero-field splitting (ZFS) of the NVs as a function of the temperature we read from our sensor and determined the temperature dependence of the ZFS to $\delta D = \SI{-92 \pm 1}{\kilo\hertz\per\kelvin}$.
    All temperature values given in this paper are calculated from a measured resonance or ZFS at the increased temperature with respect to the resonance or ZFS at room temperature. 
    Our temperature sensor determines room temperature to be \SI{19.77 \pm 0.03}{\degreeCelsius}.
    When mounting the diamond sample in the sample holder, a systematic error arises due to a slightly varying distance of the Peltier elements to the diamond chip. 
    We, therefore, calculate a statistical fit error from Lorentzian fitting to the NV spin resonances in the spectra and determine the temperature by calculating the change in the ZFS. 
    We then combine this error and the difference of our calculated temperature from ODMR spectroscopy to the mean temperature determined by the temperature sensor to the temperature errors in the main text.

    \subsection{Temperature dependence of $T_1$}
\begin{figure}[h]
    	\includegraphics[width=86mm]{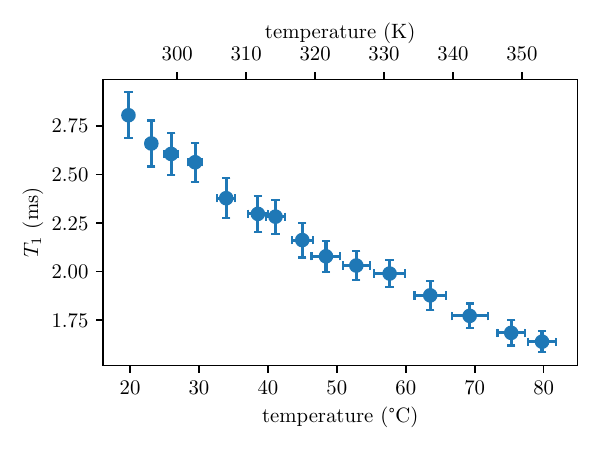}
            \caption{\label{fig:T1_temperature_dep}
          	$T_1$ of the NVs in the diamond chip as a function of the temperature, recorded with widefield relaxometry.
           }
    \end{figure}

    Since we perform temperature-dependent measurements in our study, we independently investigate the $T_1$ time as a function of the temperature of the NV-center layer in our diamond chip without an external magnetic field being applied. 
    The experiments were performed as described in the method section in widefield configuration, with \SI{30}{} accumulated exposures and $N = 1400$ repetitions.
    As opposed to the evaluation for the SCO-layer samples, we do not evaluate the $T_1$ curves per pixel since we assume a uniform $T_1$ time throughout the diamond sample.
    The mean fluorescence-difference signal as a function of $\tau$ is obtained by forming the sum of fluorescence counts of \qtyproduct[product-units=single]{600 x 600}{\px \squared} and subtracting the signal with an MW pulse applied from the signal without the MW pulse.
    We fit a biexponential function to the measurement data described in the main text.
    The results for $T_1$ are shown in Fig.~\ref{fig:T1_temperature_dep}.

    In addition to this study, we performed a similar analysis of the $T_1$ time with an external magnetic field applied. 
    We note that the $T_1$ time at room temperature coincides with the $T_1$ time measured at zero field, and the temperature dependence shows a similar behavior. 
    
    \subsection{Raman spectra}
    We record Raman spectra of the SCO thin layers deposited on the diamond.
    For \textbf{SCO I}, the sample described in the main text was investigated by Raman spectroscopy. 
    In the case of \textbf{SCO II}, we noted that the thin layers presented in the main text do not yield reasonable Raman spectra due to insufficient signal intensity. 
    We prepared a slightly thicker layer of \textbf{SCO II} using solution 2 and successively drop-cast \SI{20}{\micro \liter} onto the diamond chip and remove the solvent (see Methods section).
    The Raman spectra of \textbf{SCO II} below stem from this sample. 
    We expect the thinner layers presented in the main text to have equal spin-transition temperatures.
    Since the samples of thicker SCO layer detach from the diamond after heating and cooling, they are inaccessible to NV-$T_1$ measurements in our configuration.

\begin{figure}[h]
    	\includegraphics[width=86mm]{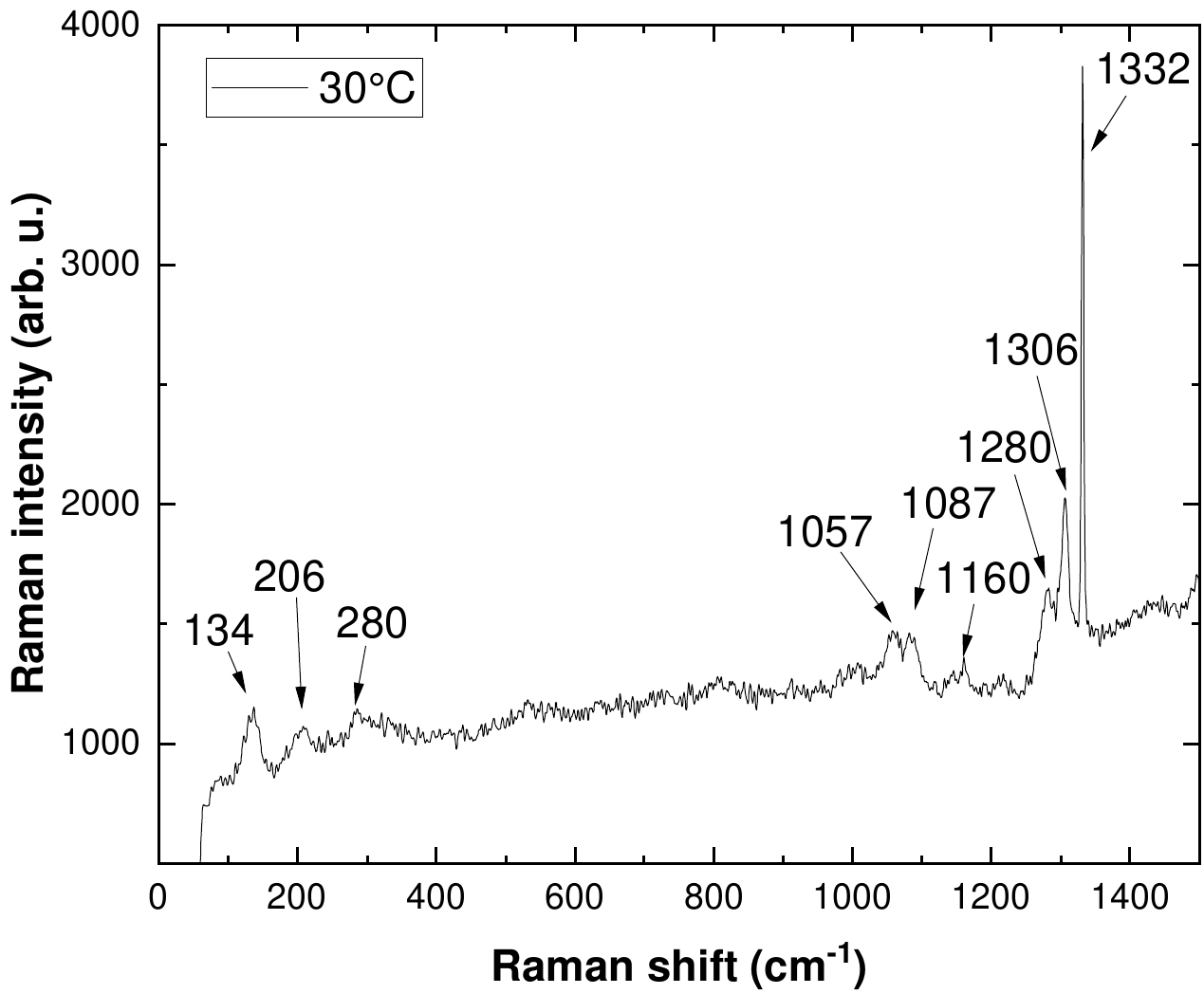}
            \caption{\label{fig:Raman_BF4_30C}
          	Raman spectrum for the thin-layer sample of \textbf{SCO~I} applied on diamond at \SI{30}{\degreeCelsius}.
           }
        \end{figure}

\begin{figure}[h]
    	\includegraphics[width=86mm]{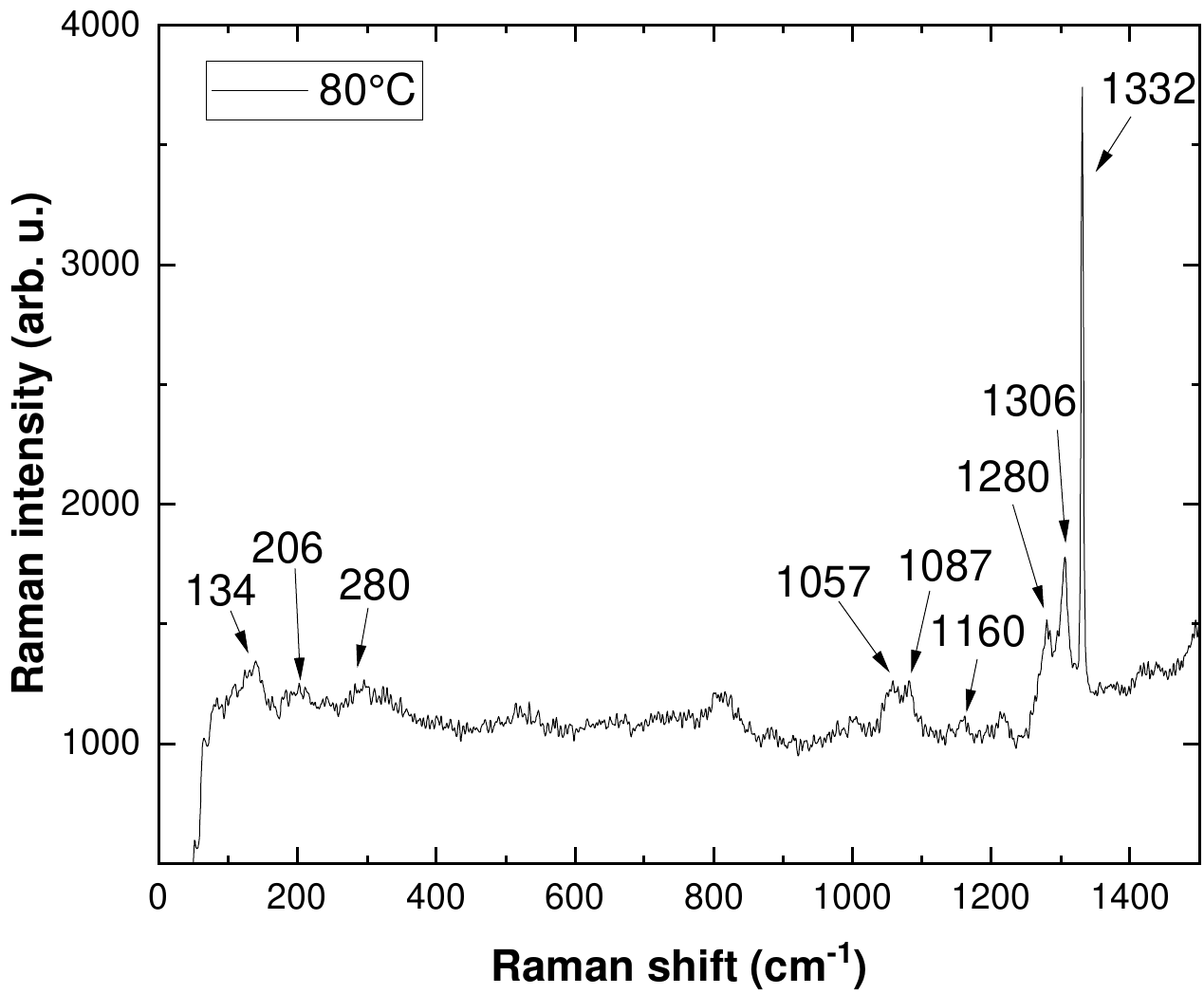}
            \caption{\label{fig:Raman_BF4_80C}
          	Raman spectrum for the thin-layer sample of \textbf{SCO~I} applied on diamond at \SI{80}{\degreeCelsius} in the heating branch.
           }
        \end{figure}

\begin{figure}[h]
    	\includegraphics[width=86mm]{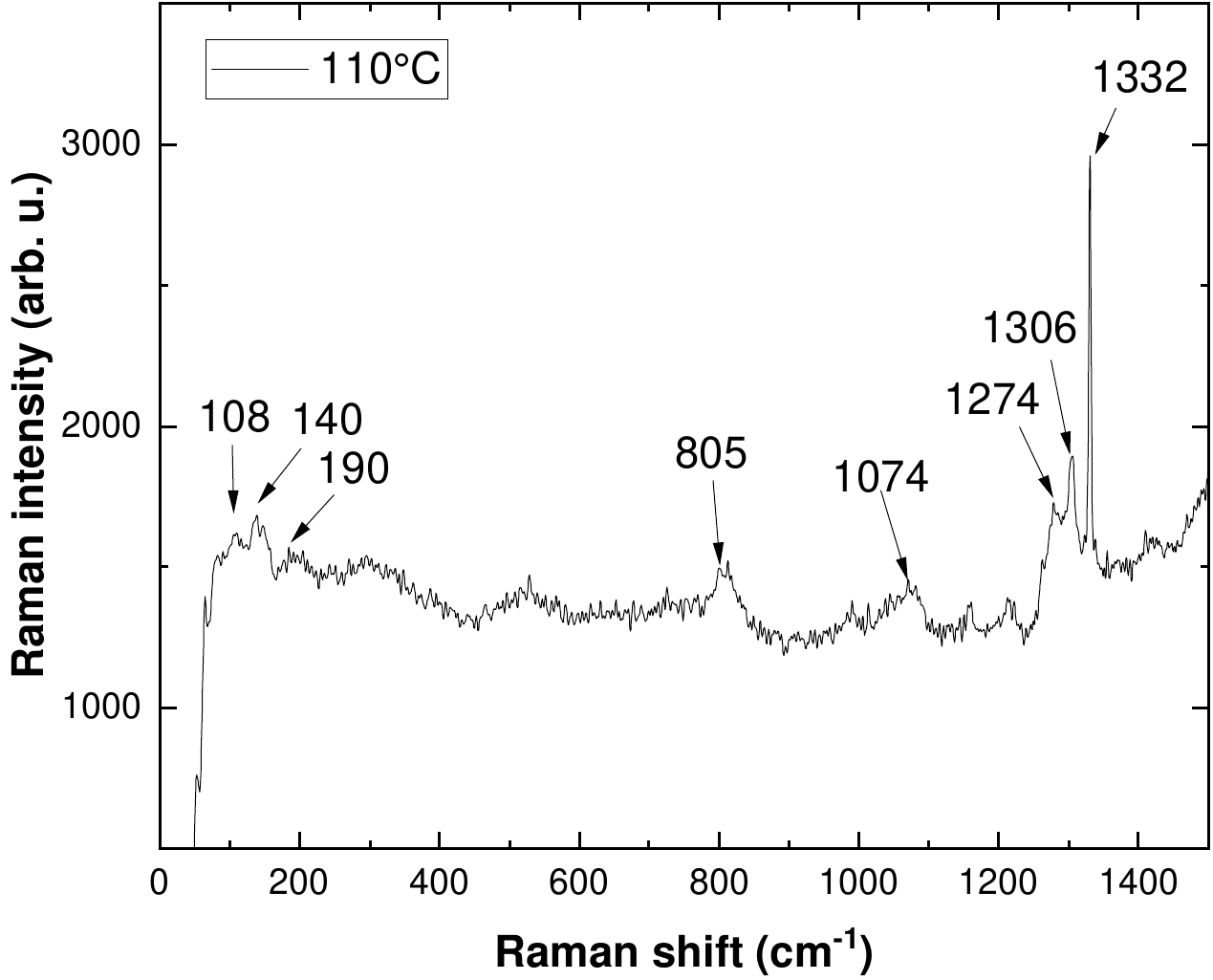}
            \caption{\label{fig:Raman_BF4_110C}
          	Raman spectrum for the thin-layer sample of \textbf{SCO~I} applied on diamond at \SI{110}{\degreeCelsius}.
           }
        \end{figure}

\begin{figure}[h]
    	\includegraphics[width=86mm]{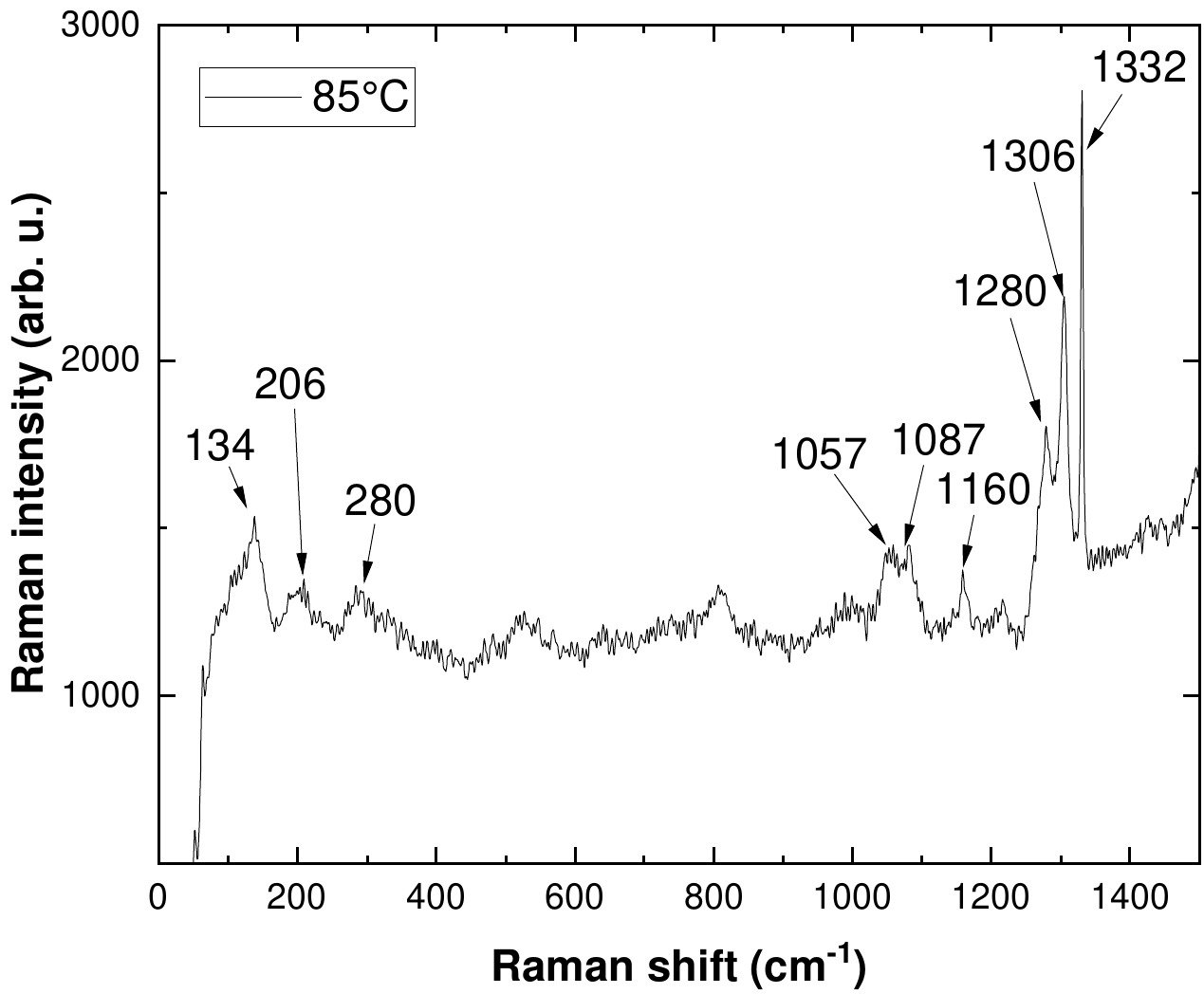}
            \caption{\label{fig:Raman_BF4_85C}
          	Raman spectrum for the thin-layer sample of \textbf{SCO~I} applied on diamond at \SI{85}{\degreeCelsius} in the cooling branch.
           }
        \end{figure}

   The Raman spectrum of the thin-layer of \textbf{SCO I} applied on diamond is depicted in Figs.~\ref{fig:Raman_BF4_30C} to \ref{fig:Raman_BF4_85C} at \SI{30}{\degreeCelsius}, \SI{80}{\degreeCelsius}, \SI{110}{\degreeCelsius}, and \SI{85}{\degreeCelsius}, respectively. 
   At \SI{30}{\degreeCelsius}, where the sample is predominantly in the LS state, there are three Raman bands at \SI{134}{}, \SI{206}{}, and \SI{280}{\per \centi\meter}. 
   As these bands are not observed in the Raman spectrum at \SI{110}{\degreeCelsius}, we conclude that these bands represent LS marker bands. 
   Accordingly, the bands at \SI{108}{}, \SI{140}{}, and \SI{190}{\per \centi\meter} represent HS marker bands. Further marker bands in LS state can be noted at \SI{1057}{}, \SI{1087}{}, \SI{1160}{}, \SI{1280}{}, \SI{1306}{}, and \SI{1332}{\per \centi\meter} as well as \SI{805}{}, \SI{1074}{}, \SI{1274}{}, and \SI{1306}{\per \centi\meter} in the HS state. 
   The intense peak at \SI{1332}{\per \centi\meter} is characteristic for the diamond substrate \cite{Gorelik.2020_S}, 
   while the other modes are typical for the thin-layer of the \textbf{SCO I} complex \cite{Urakawa.2011_S}. 
   Based on the spectra, it can be concluded that the sample is predominantly in the LS state at the temperatures used for the determination of the $T_1$ times.

\begin{figure}[h]
    	\includegraphics[width=86mm]{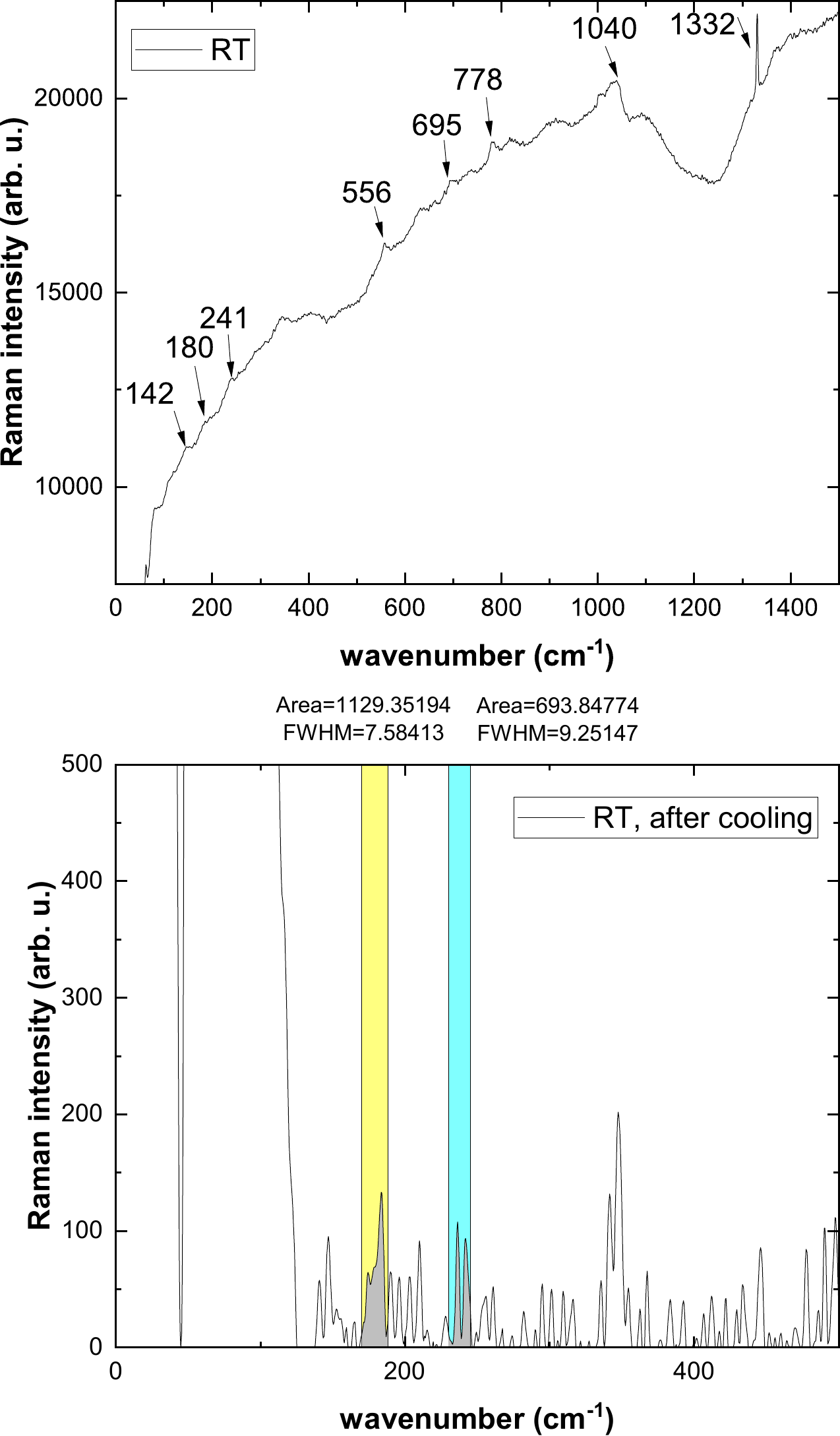}
            \caption{\label{fig:Raman_MSFSO4_RT}
          	Full- and background-corrected Raman spectrum for the thin-layer sample of \textbf{SCO~II} applied on diamond at room temperature in the heating branch.
           }
        \end{figure}

\begin{figure}[h]
    	\includegraphics[width=86mm]{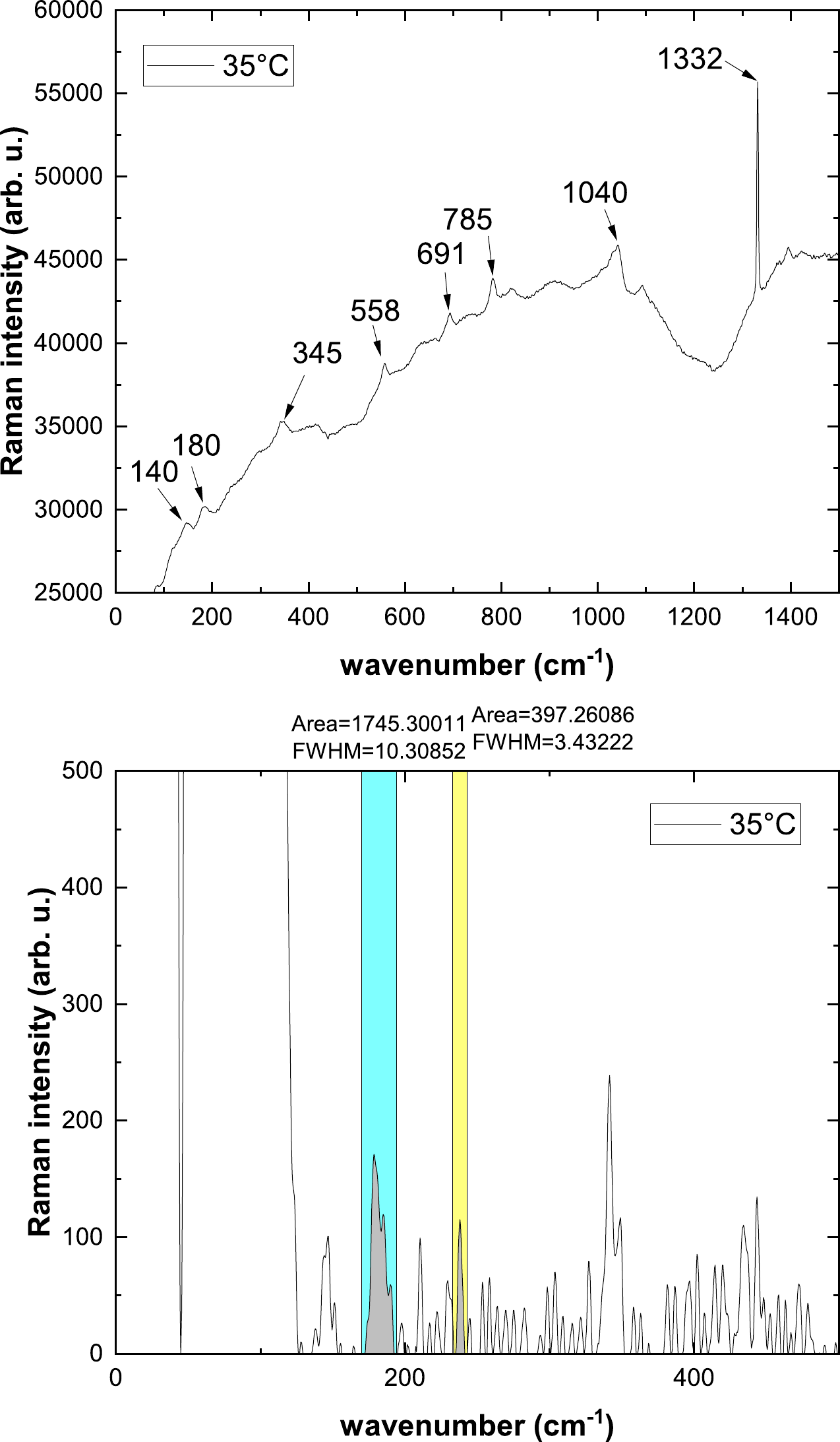}
            \caption{\label{fig:Raman_MSFSO4_35C}
          	Full- and background-corrected Raman spectrum for the thin-layer sample of \textbf{SCO~II} applied on diamond at \SI{35}{\degreeCelsius}.
           }
        \end{figure}

\begin{figure}[h]
    	\includegraphics[width=86mm]{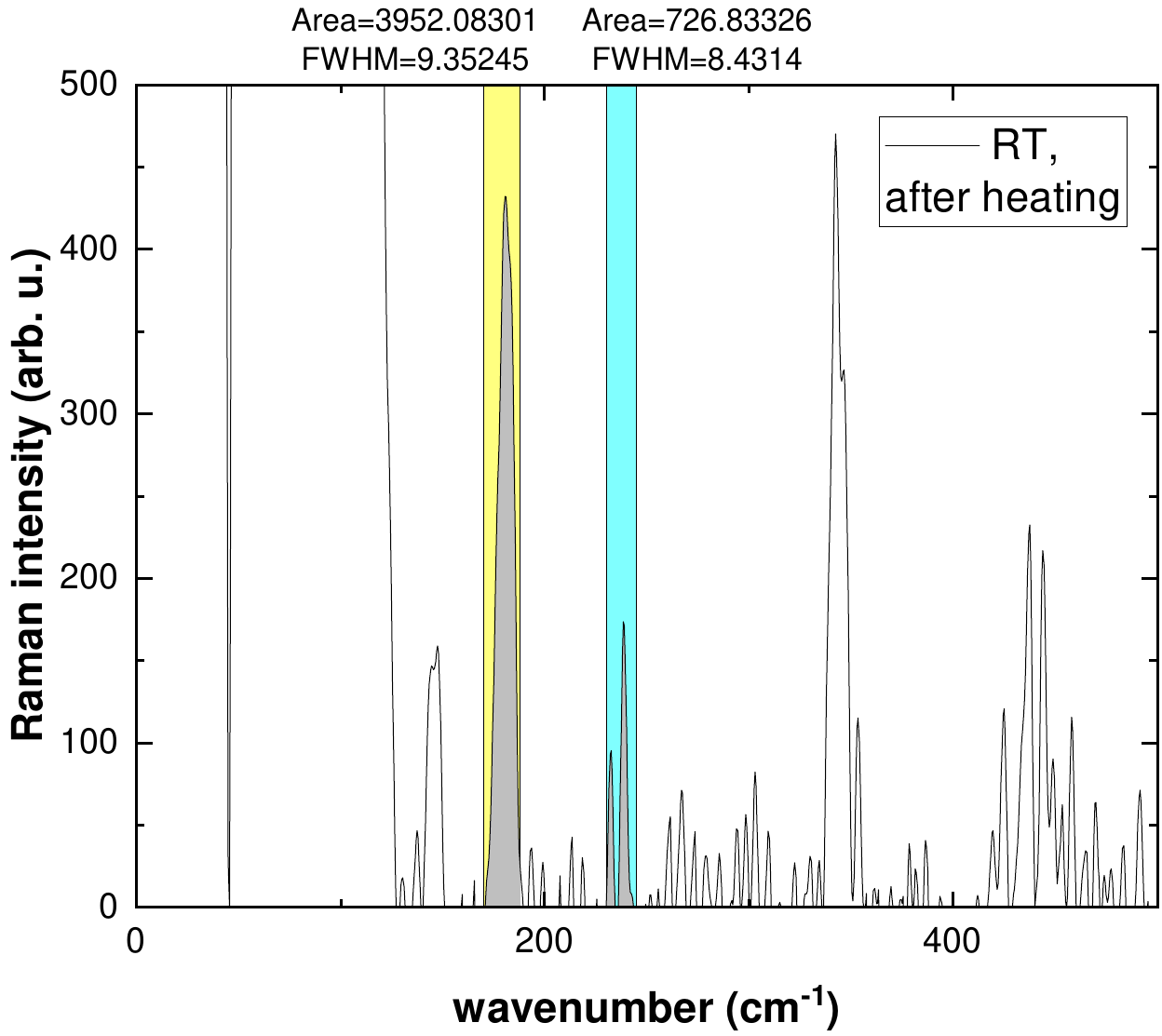}
            \caption{\label{fig:Raman_MSFSO4_RT_down}
          	Raman spectrum for the thin-layer sample of \textbf{SCO~II} applied on diamond at room temperature in the cooling branch.
           }
        \end{figure}

   Furthermore, the Raman spectra of the thin-layer of \textbf{SCO II} deposited on diamond at room temperature and at \SI{35}{\degreeCelsius} upon heating are shown in Fig.~\ref{fig:Raman_MSFSO4_RT} and Fig.~\ref{fig:Raman_MSFSO4_35C}. 
   For comparison of the heating and cooling branch of the SCO transition, a further Raman spectrum at room temperature upon cooling from \SI{35}{\degreeCelsius} is displayed in Fig.~\ref{fig:Raman_MSFSO4_RT_down}. 
   On the one hand, the Raman spectrum at room temperature in the heating branch shows vibrations at \SI{142}{}, \SI{180}{}, \SI{241}{}, \SI{556}{}, \SI{695}{}, \SI{778}{}, \SI{1040}{} and \SI{1332}{\per \centi\meter}. 
   On the other hand, the Raman spectrum at \SI{35}{\degreeCelsius} reveals bands at \SI{140}{}, \SI{180}{}, \SI{345}{}, \SI{558}{}, \SI{691}{}, \SI{785}{}, \SI{1040}{} and \SI{1332}{\per \centi\meter}. 
   The vibration at \SI{1332}{\per \centi\meter} can be attributed to the diamond substrate while all other vibrations are typical for the thin-layer of \textbf{SCO II} \cite{Hochdorffer.2023_S}. 
   As the vibrations around \SI{241}{\per \centi\meter} disappear upon heating of the sample, it is concluded that these vibrations are characteristic LS marker bands. 
   Similarly, the vibrations at \SI{140}{\per \centi\meter} and \SI{180}{\per \centi\meter} represent HS marker bands. 
   Therefore, it can be determined that the sample is in a mixed HS and LS state at room temperature and in pure HS state at \SI{35}{\degreeCelsius}. 
   In order to determine the ratio of HS and LS at the different temperatures, the areas under the \SI{180}{\per \centi\meter} and \SI{240}{\per \centi\meter} vibrations were calculated and divided by the sum of both areas. 
   As all three spectra display a noticeable background and noise which are caused by all optical components of the spectrometer, the spectra were corrected using the shape correction tool integrated into the Raman spectrometer. 
   The evaluation of the areas yields \SI{62}{\percent} HS and \SI{38}{\percent} at room temperature, \SI{82}{\percent} HS and \SI{18}{\percent} LS at \SI{35}{\degreeCelsius} (the LS component occurs due to leftover noise after the shape correction at \SI{240}{\per \centi\meter}) and \SI{84}{\percent} HS and \SI{16}{\percent} LS at room temperature upon cooling. 
   Therefore, a considerable difference for the Raman spectra at room temperature can be concluded in dependence of the heating or cooling branch of the hysteresis.

   \subsection{Supporting graphs and images}
\begin{figure*}[]
    \includegraphics[width=172mm]{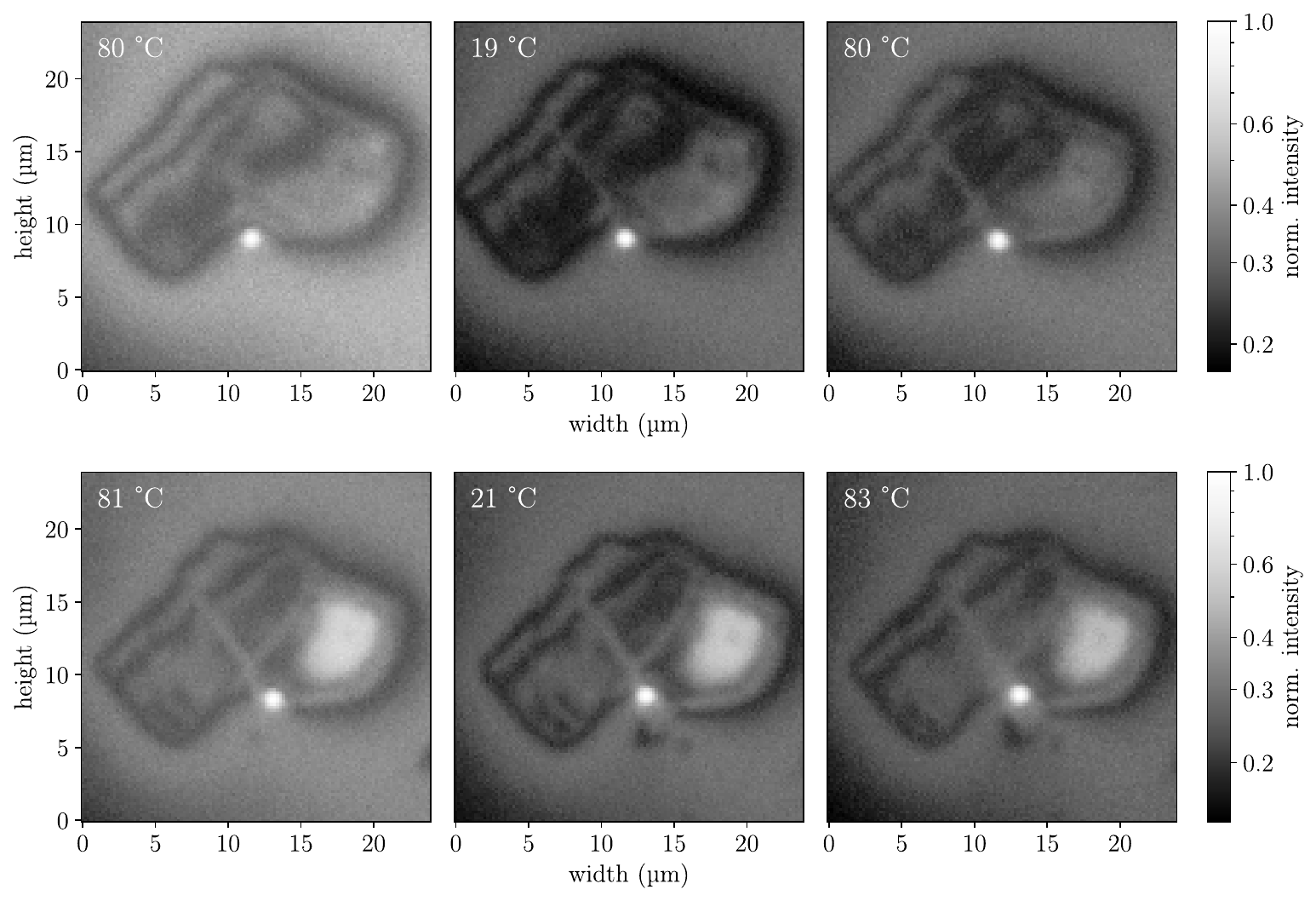}
        \caption{\label{fig:BF4_whitelight}
        LED images of the thin-layer sample of \textbf{SCO~I} for the two subsequent cycles as shown in the main text for the temperatures as given in the images.
        The images were recorded in their order of appearance at \SI{80 \pm 2}{\degreeCelsius}, \SI{19 \pm 1}{\degreeCelsius}, and \SI{80 \pm 2}{\degreeCelsius} in the first cycle (first row).
        In the second cycle, the images were recorded at \SI{81 \pm 2}{\degreeCelsius}, \SI{21 \pm 1}{\degreeCelsius}, and \SI{83 \pm 1}{\degreeCelsius} (second row).
        The exact laser position is marked as a bright spot (filtered NV fluorescence). 
        A structural change in the sample structure is visible, which is most likely caused by the heating and cooling of the sample.
       }
    \end{figure*}

\begin{figure*}[]
    \includegraphics[width=172mm]{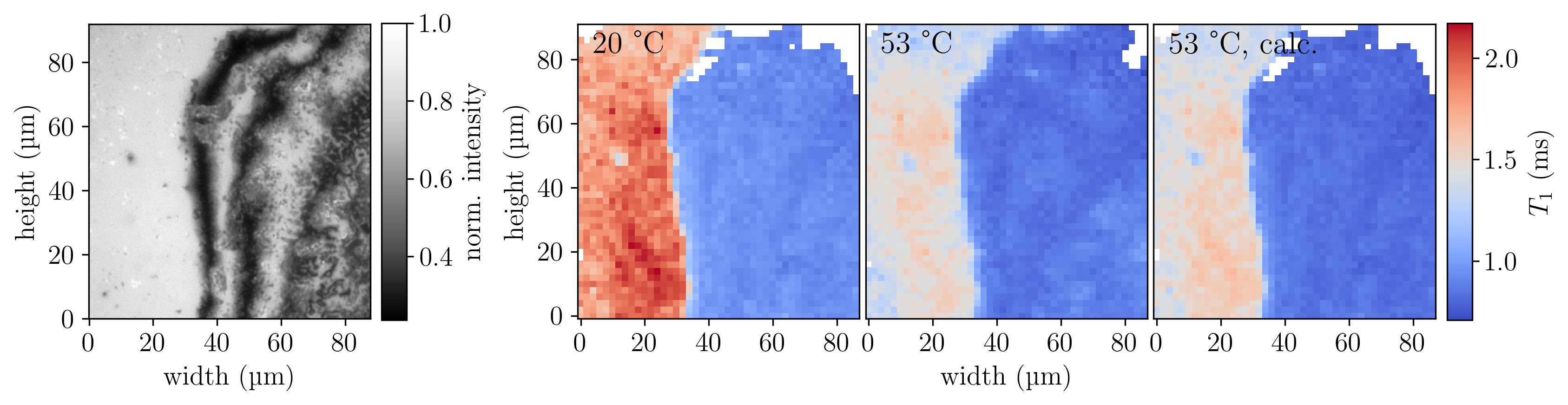}
        \caption{\label{fig:msf_so4_temperature_estimation}
        LED image and $T_1$ measurements for the thin-layer sample of \textbf{SCO~II}, sample 1, for different temperatures. 
        The $T_1$ measurements at \SI{20 \pm 1}{\degreeCelsius} and at \SI{53 \pm 2}{\degreeCelsius} are the same as in Fig.~\ref{fig:MSFSO4thin}.
        We separate $T_1^{\mathrm{SCO}}$ from $T_1^{\prime}$ at \SI{20}{\degreeCelsius} using Eq.~\ref{eq:T1_SCO} and add it to $T_1^{\prime}$ at \SI{53}{\degreeCelsius}.
        This way, we obtain the third $T_1$ map, which describes a calculated $T_1$ time at \SI{53}{\degreeCelsius} with the NV-$T_1$ temperature dependence, but an unchanged SCO influence.
        Since the difference of the measured and calculated $T_1$ maps at \SI{53}{\degreeCelsius} is in the order of the $T_1$ uncertainties, we exclude influences of an SCO spin switching on the $T_1$ times.
        }
    \end{figure*}

\begin{figure*}[]
    \includegraphics[width=172mm]{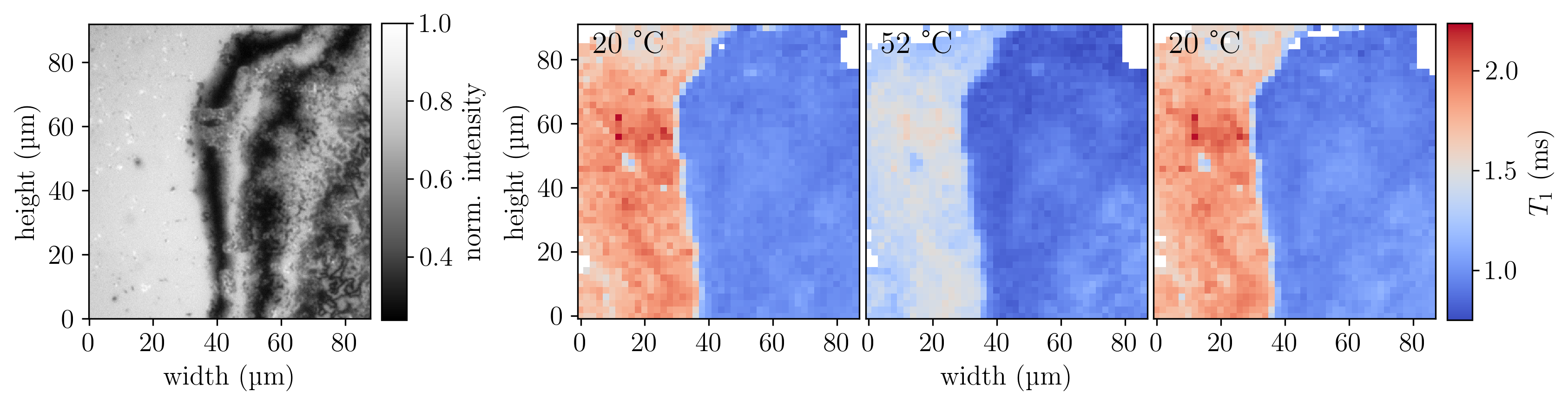}
        \caption{\label{fig:cycle_2_sco_msf_so4}
       LED image and $T_1$ map for the same region at different temperatures for the thin-layer sample 1 of \textbf{SCO~II}.
       The results are from a second heating and cooling cycle as presented in the main text, see Fig.~\ref{fig:MSFSO4thin}.
       The $T_1$ maps were recorded in order as presented.
       After cooling the sample to \SI{-20}{\degreeCelsius}, we record the $T_1$ map at \SI{20 \pm 1}{\degreeCelsius} in the heating branch. 
        Then, the $T_1$ times are measured at \SI{52 \pm 2}{\degreeCelsius}.
        Lastly, the sample is cooled back to \SI{20 \pm 1}{\degreeCelsius} to record the $T_1$ times in the cooling branch.
       }
    \end{figure*}

\begin{figure*}[]
    \includegraphics[width=172mm]{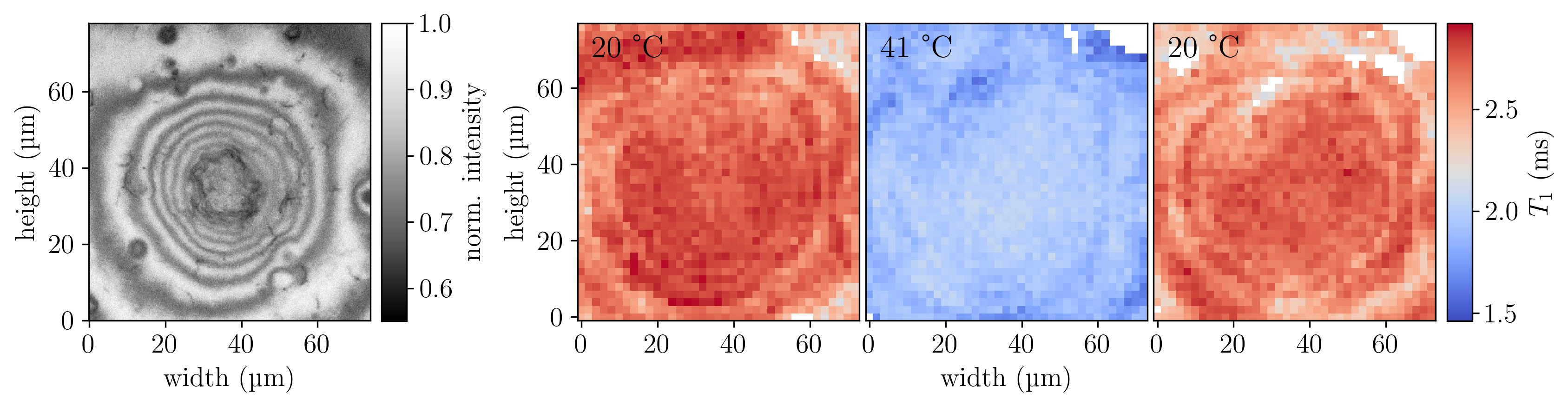}
        \caption{\label{fig:pos_sco_msf_so4}
       LED image and $T_1$ map for the same region at different temperatures for the thin-layer sample 2 of \textbf{SCO~II}.
       The results are from the same sample presented in Fig.~\ref{fig:MSFSO4thick} but at a different position.
       The $T_1$ maps were recorded in order as presented.
       After cooling the sample to \SI{-20}{\degreeCelsius}, we record the $T_1$ map at \SI{20 \pm 1}{\degreeCelsius} in the heating branch. 
        Then, the $T_1$ times are measured at \SI{41 \pm 2}{\degreeCelsius}.
        Lastly, the sample is cooled back to \SI{20 \pm 1}{\degreeCelsius} to record the $T_1$ times in the cooling branch.
       }
    \end{figure*}

\begin{figure}[]
    	\includegraphics[width=86mm]{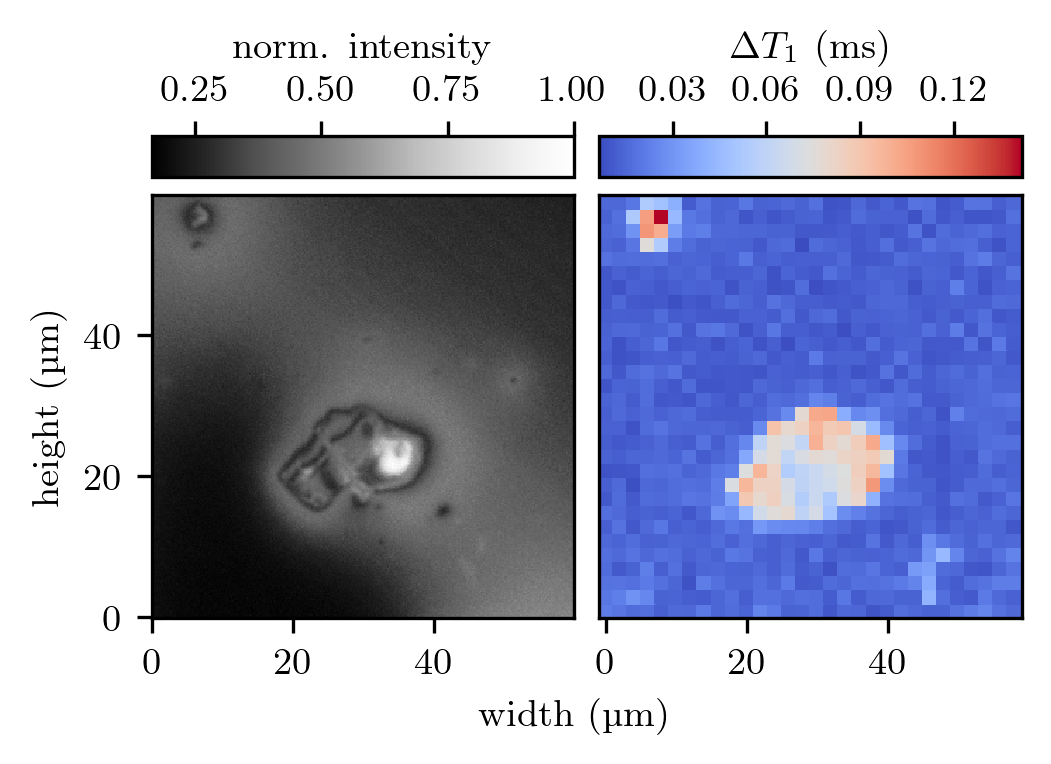}
            \caption{\label{fig:T1_BF4_errors}
          	LED image and values for $\Delta T_1$ of the thin-layer sample of \textbf{SCO~I}, supporting Fig.~\ref{fig:BF4}.
           }
    \end{figure}

\begin{figure*}[]
    \includegraphics[width=172mm]{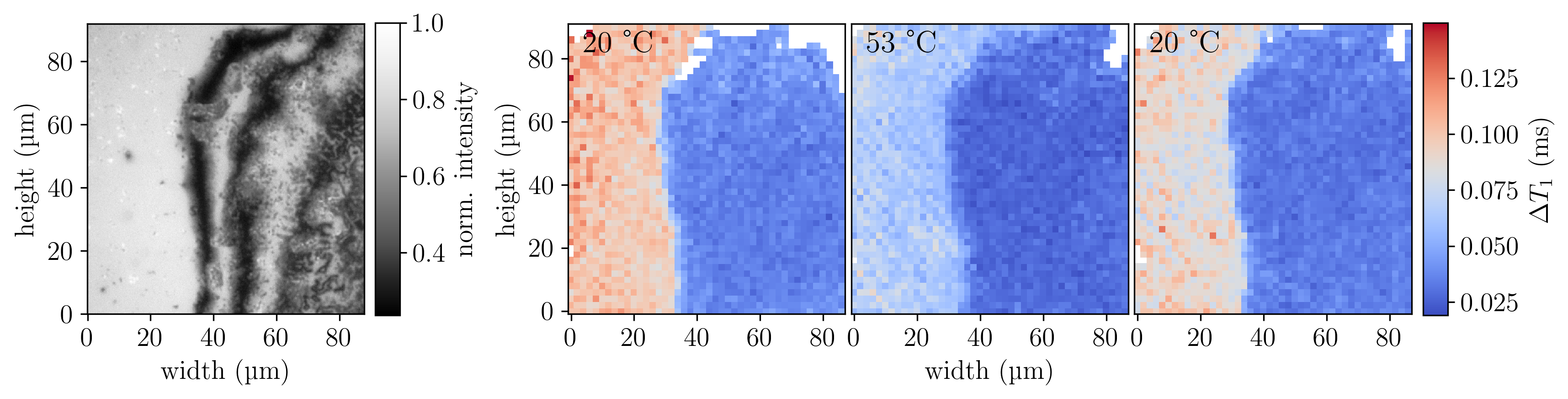}
        \caption{\label{fig:MSF_SO4_errors_cylce_1}
       LED image and values for $\Delta T_1$ for different temperatures of the thin-layer sample 1 of \textbf{SCO~II} in the first cycle, supporting Fig.~\ref{fig:MSFSO4thin}.
       }
    \end{figure*}

\begin{figure*}[]
    \includegraphics[width=172mm]{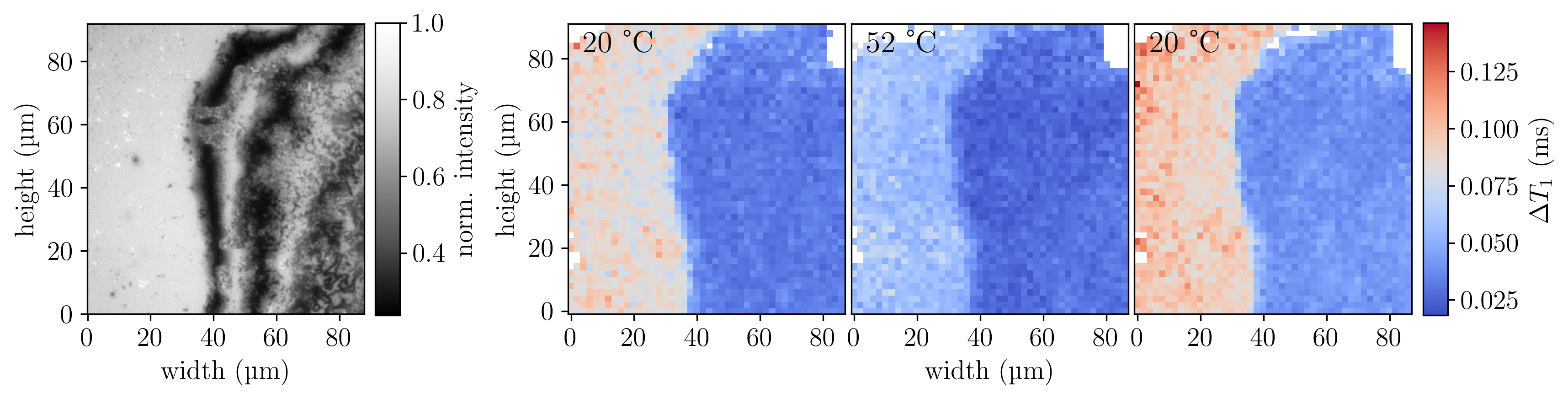}
        \caption{\label{fig:MSFSO4thin_err}
       LED image and values for $\Delta T_1$ for different temperatures of the thin-layer sample 1 of \textbf{SCO~II} in the second cycle, supporting Fig.~\ref{fig:cycle_2_sco_msf_so4}.
       }
    \end{figure*}

\begin{figure*}[]
    \includegraphics[width=172mm]{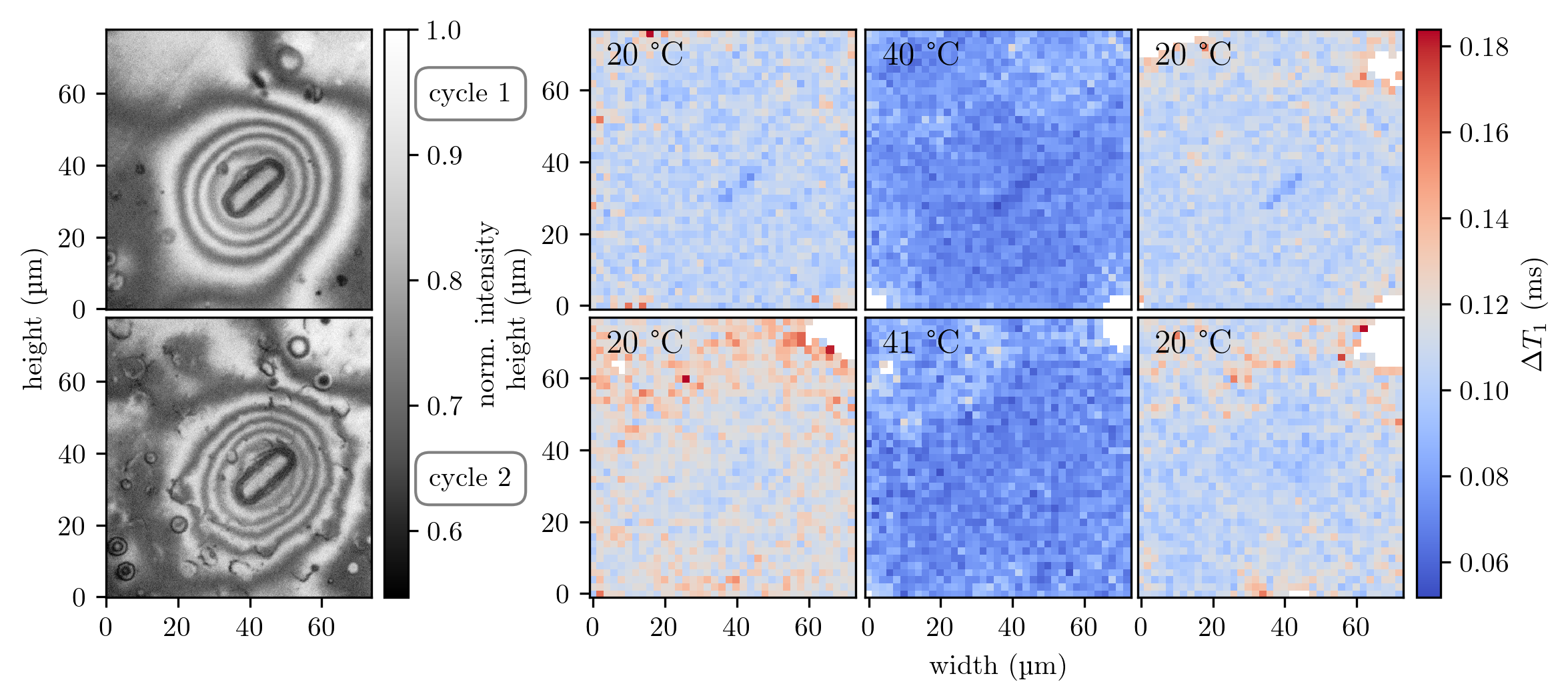}
        \caption{\label{fig:MSF_SO4_errors_rectangle}
        LED image and values for $\Delta T_1$ for different temperatures of the thin-layer sample 2 of \textbf{SCO~II} for two subsequent cycles, supporting Fig.~\ref{fig:MSFSO4thick}.
       }
    \end{figure*}

\begin{figure*}[]
    \includegraphics[width=172mm]{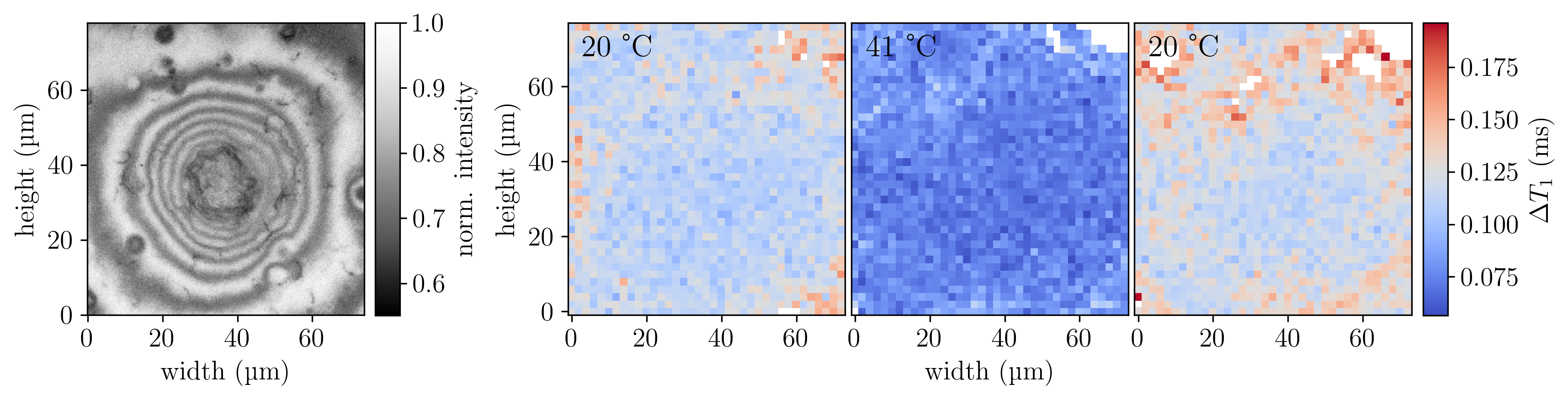}
        \caption{\label{fig:MSF_SO4_errors_pos2}
       LED image and values for $\Delta T_1$ for different temperatures of the thin-layer sample 2 of \textbf{SCO~II} in a second position, supporting Fig.~\ref{fig:pos_sco_msf_so4}.
       }
    \end{figure*}

\begin{figure*}[]
    \includegraphics[width=172mm]{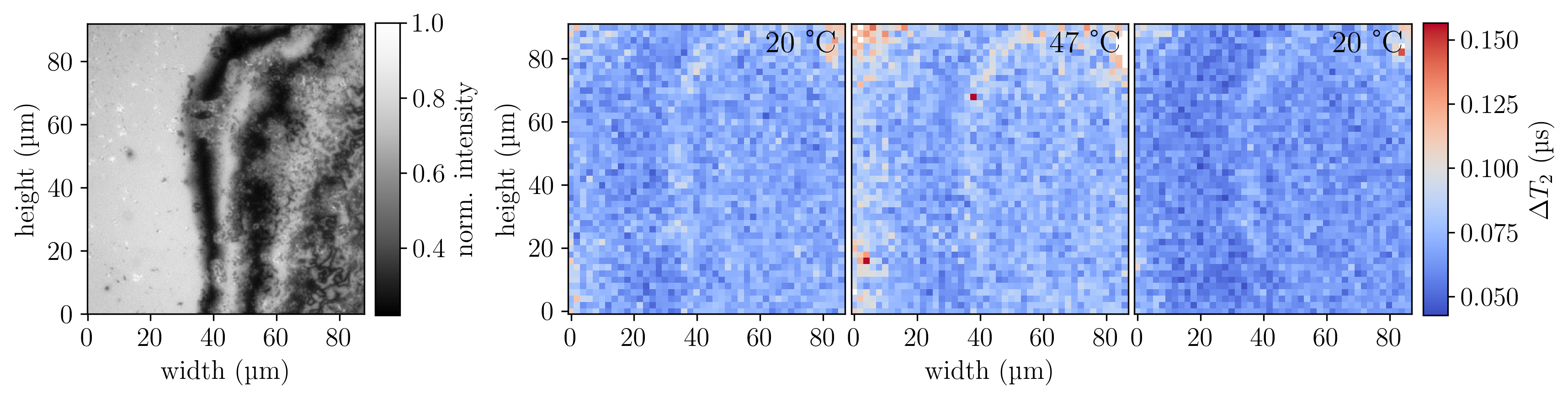}
        \caption{\label{fig:Hahn_errors}
       LED image and values for $\Delta T_2$ for different temperatures of the thin-layer sample 1 of \textbf{SCO~II}, supporting Fig.~\ref{fig:MSFSO4Hahn}.
       }
    \end{figure*}

\begin{figure}[]
    	\includegraphics[width=86mm]{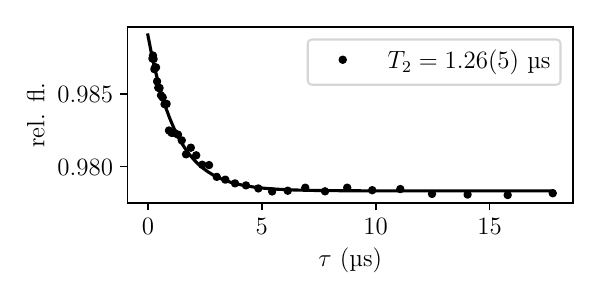}
            \caption{\label{fig:hahn}
          	Results of the $T_2$ measurement (Hahn echo) of the NV center in the diamond chip without any SCO complexes applied.
            We measured as described previously.
            Since we assume the $T_2$ time to be equal for the entire diamond chip, we do not evaluate the relative fluorescence for each pixel. 
            The relative fluorescence signal as a function of $\tau$ is obtained by calculating the sum of fluorescence counts of \qtyproduct[product-units=single]{600 x 600}{\px \squared} and forming the quotient of the signal with the MW pulses applied and the signal without the MW pulses. 
            The solid line is a fit curve to the measurement points described in the main text.
            The data was recorded in an external magnetic field of amplitude in the order of \SI{11}{\milli \tesla}.
            The NV centers of one crystal orientation were probed at $\SI{\approx 2817}{\mega \hertz}$.
           }
    \end{figure}

    In this section, we plot additional graphs and images. 
    These include LED images of \textbf{SCO~I} for two subsequent heating and cooling cycles, see Fig.~\ref{fig:BF4_whitelight}.
    In Fig.~\ref{fig:msf_so4_temperature_estimation}, we estimate the combined effect of a temperature increase and the presence of paramagnetic SCO complexes on the NV centers' $T_1$ time for a $T_1$ map of \textbf{SCO~II}.
    Further, additional $T_1$ maps for measurements on \textbf{SCO~II} are shown in Figs.~\ref{fig:cycle_2_sco_msf_so4} and \ref{fig:pos_sco_msf_so4}.
    The $\Delta T_1$ maps, which include the statistical fit errors of all $T_1$ matrices shown in this paper, are displayed in Figs.~\ref{fig:T1_BF4_errors}, \ref{fig:MSF_SO4_errors_cylce_1}, \ref{fig:MSFSO4thin_err}, \ref{fig:MSF_SO4_errors_rectangle}, and \ref{fig:MSF_SO4_errors_pos2}.
    In Fig.~\ref{fig:Hahn_errors}, the $\Delta T_2$ maps are shown.
    Lastly, a $T_2$ measurement for the NVs in the clean diamond (no SCO complexes applied) is shown in Fig.~\ref{fig:hahn}.

    \FloatBarrier
  
    %

\end{document}